\documentstyle[10pt]{article}
\title{Two dimensional lattice gauge theory based on a quantum group}
\author{E.Buffenoir\thanks{e-mail: buffenoi@orphee.polytechnique.fr},
Ph.Roche\thanks{e-mail: roche@orphee.polytechnique.fr}\cr
Centre de Physique Theorique de L' Ecole Polytechnique
\thanks{Laboratoire Propre du CNRS UPR 14}\cr
91128 Palaiseau cedex\cr
France}

\date{\today}
\begin{document}
\maketitle

\begin{abstract}
In this article we   analyze  a two dimensional
lattice gauge theory based on a quantum group. The algebra generated by
gauge fields is the lattice algebra
introduced recently by A.Yu.Alekseev, H.Grosse and V.Schomerus   in
\cite{AGS}.
We define and study Wilson loops.
This
theory  is quasi-topological as in the classical case, which allows us to
compute the
correlation  functions of this  theory on an arbitrary surface.

Preprint: CPTH A302-05/94
\end{abstract}

\def\be{\begin{equation}}
\def\ee{\end{equation}}
\def\bea{\begin{eqnarray}}
\def\eea{\end{eqnarray}}

\def\Rab{{\buildrel \alpha\beta \over R}\!}
\def\Raa{\buildrel \alpha\alpha \over R}
\def\Rabara{\buildrel {\bar\alpha}\alpha \over R}
\def\Rabbar{\buildrel {\bar\alpha}{\bar\beta} \over R}
\def\Rff{\buildrel ff \over R}

\def\Rba{\buildrel \beta\alpha \over R}
\def\End{\rm End}
\def\Hom{{\rm Hom}_A}
\def\ea{\buildrel\alpha \over e}
\def\eb{\buildrel\beta \over e}
\def\ec{\buildrel\gamma \over e}
\def\ga{\buildrel\alpha \over g}
\def\gabar{\buildrel{\bar\alpha} \over g}

\def\gb{\buildrel\beta \over g}
\def\gc{\buildrel\gamma \over g}
\def\gf{\buildrel f \over g}
\def\ua{{\buildrel\alpha \over U}\!}
\def\uo{{\buildrel 0 \over U}\!}
\def\uabar{{\buildrel{\bar\alpha} \over U}\!}
\def\ubbar{{\buildrel{\bar\beta} \over U}\!}
\def\ugbar{{\buildrel{\bar\gamma} \over U}\!}
\def\ub{{\buildrel\beta \over U}\!}
\def\uc{{\buildrel\gamma \over U}\!}
\def\Int{\mbox{Int}}
\def\sua{\buildrel\alpha\over u}
\def\mua{\buildrel\alpha\over \mu}
\def\mub{\buildrel\beta\over \mu}
\def\muc{\buildrel\gamma\over \mu}
\def\subeta{\buildrel\beta\over u}
\def\sug{\buildrel\gamma\over u}
\def\aa{\buildrel\alpha \over a}
\def\ab{\buildrel\beta \over a}
\def\ac{\buildrel\gamma \over a}
\def\ba{\buildrel\alpha \over b}
\def\bb{\buildrel\beta \over b}
\def\bc{\buildrel\gamma \over b}
\def\ca{\buildrel\alpha \over c}
\def\cb{\buildrel\beta \over c}
\def\cg{\buildrel\gamma \over c}
\def\da{\buildrel\alpha \over d}
\def\db{\buildrel\beta \over d}
\def\dc{\buildrel\gamma \over d}
\def\ea{\buildrel\alpha \over e}
\def\eb{\buildrel\beta \over e}
\def\ec{\buildrel\gamma \over e}
\def\xa{\buildrel\alpha \over x}
\def\xb{\buildrel\beta \over x}
\def\xc{\buildrel\gamma \over x}
\def\ya{\buildrel\alpha \over y}
\def\yb{\buildrel\beta \over y}
\def\yc{\buildrel\gamma \over y}
\def\za{\buildrel\alpha \over z}
\def\zb{\buildrel\beta \over z}
\def\zc{\buildrel\gamma \over z}
\def\chia#1{\buildrel\alpha_{#1}\over \chi}
\def\chib#1{\buildrel\beta_{#1}\over \chi}
\def\Ua#1{{\buildrel\alpha_{#1} \over U}}
\def\Ub#1{{\buildrel\beta_{#1} \over U}}

\def\CGphi#1#2#3#4#5#6#7{\left(\matrix{#2 & #4\cr
#1 & #3 \cr}\right\vert
\left.\matrix{ #5\cr
#6 \cr}\right)_{#7}^{\phi}}

\def\CGpsi#1#2#3#4#5#6#7{\left(\matrix{#6 \cr
#5  \cr}\right\vert
\left.\matrix{ #1&#3 \cr
#2& #4  \cr}\right)_{#7}^{\psi}}
\newtheorem{definition}{Definition}
\newtheorem{proposition}{Proposition}
\def\bd{\begin{definition}}
\def\ed{\end{definition}}
\def\bp{\begin{proposition}}
\def\ep{\end{proposition}}

\section{Introduction}
 Quantum groups appeared in the mid-eighties  as hidden algebraic
structures generalizing the notion of group symmetries in integrable
systems \cite{Ji}. There are now different definitions of quantum groups
which
include  the local point of view (deformation of the Lie algebra) as well
as the
global point of view (deformation of the algebra of continuous functions
on a Lie group).

The latter provides examples  of quantum geometry, and the
ordinary tools of differential geometry on Lie groups can be successfully
defined and used to study, for example, harmonic analysis on quantum
groups \cite{Wo}. This success has encouraged people to apply these tools
to
build examples of quantum geometry where the notion of group symmetry is
essential: quantum vector spaces, quantum homogeneous spaces, quantum
principal fiber bundles \cite{BM}. It is then tempting to hope that quantum
groups can be used in a much  broader area than just integrable models,
and could give, as an example,  a Yang Mills type theory associated to a
quantum group, leading hopefully to  new Physics. There has been quite
a lot of work dealing with q-deformed Yang Mills theory with a base
space being a classical space or a quantum space. These works only
deal with the study of what could be called classical configurations of the
gauge fields, but do not study the path integral on the space of
connections. The
work of \cite{BM} although perfectly coherent for classical configurations
requires a finer analysis when it studies path integrals, because the gauge
fields, in their work, are living in the deformation of the envelopping
algebra
and it is not at all obvious to define the path integral on the space of
connections taking values not in the Lie algebra but in the envelopping
algebra
and to show that when the parameter q goes to one, one recovers ordinary
quantum
Yang Mills.

Another way to cope with the problem of summing over all the configurations
is
to use a lattice regularization of the theory \cite{Cr}. This is what has
been
first described by D.V.Boulatov  \cite{Bo}.   He studied there
the  q-deformation of Wilson lattice formulation of gauge theories in
dimension
two and three and computed the lattice partition function of these models
when q is a root of unity and when the Yang Mills coupling constant goes to
$0.$   He argues that the partition function of this q-deformed lattice
model is the  Turaev-Viro invariant in dimension 3 and
 the partition function of $(G/G)_k$ models in dimension 2.
This theory is unfortunately not consistent with the
gauge invariance and moreover we have shown (unpublished) that it
is equivalent, in the two dimensional case, to the undeformed Yang-Mills
theory.

In their recent work \cite{AGS}, A.Alekseev $\&$ al. have studied a
combinatorial quantization of the hamiltonian Chern-Simons theory. They
were
led to define an algebra of gauge fields on which acts the gauge quantum
group.
This algebra is an exchange algebra which appears to be a generalization of
the discretization of the current algebra found
by \cite{AFS}\cite{BB}. The commutation relations between different link
variables
are quantization of
those found by V.V.Fock $\&$ A.A.Rosly \cite{FR}.

In our present work we will define and study a quantum deformation of a
lattice
gauge theory on a triangulated surface. The gauge symmetry is described by
a
deformation of the algebra of continuous functions on a Lie group. The
gauge
fields living on the links generate an algebra on which coacts the gauge
symmetry.
Invariance under the gauge symmetry implies that this algebra is the
algebra of
gauge fields of  A.Alekseev, H.Grosse and V.Schomerus. Our presentation of
the gauge symmetry is dual to these authors and do not involve any
involution.
Using this point of view computations are greatly simplified.

 Wilson loops in our formalism are obtained
using the notion of quantum trace and have the usual properties: invariance
under departure
point, gauge invariance. We will use them as usual to define the Boltzmann
weights of the theory.
 These Boltzmann weights satisfy the
familiar convolution property of two-dimensional Yang Mills theory which
implies
that the theory is quasitopological, i.e physical
quantities depend only on the topology and the area of the surface. We are
able to compute
correlation functions in this model, they finally appear to be the q-analog
of the correlation functions
of ordinary two dimensional Yang Mills theory.

The present work is divided in four sections:
in the  first section of this work we recall standard properties of two
dimensional Yang Mills
theory.
In  the second section we  derive and study the algebra of gauge fields using
gauge covariance.
The third section is
devoted to the definition of Wilson loops and the study of their
commutation properties.
In the fourth section we define and study the analog of standard tools of
quantum lattice
gauge field
theory i.e Boltzmann weights, Yang-Mills measure. We compute the
correlation functions, i.e
partition functions on Riemann surfaces with punctures.

\section{Algebraic properties of ordinary lattice gauge theory}

In this part we  recall  algebraic properties of ordinary lattice gauge
theories associated to a compact group $G.$

Let $R$ be a D-dimensional lattice, $R= {\bf Z}^D $ and denote by  $V$ the
set of vertices of
$R$   and  $L$  the  set of oriented
links, i.e  the set of couple of nearest neighbour points of $R.$

A lattice gauge theory is defined by assigning to each link $l=(i,j)$
an element
$U_l$ of $G$ satisfying

\begin{equation}
U_{(i,j)}U_{(j,i)}=1.\label{uu=1}
\end{equation}

 These link variables
interact in
 a gauge invariant way. The group of gauge transformations is the
group of maps $g$ from $V$ to $G,$ which acts on the set of link
variables by:

\begin{equation}
U^g_{(i,j)}=g_i U_{(i,j)}g_j^{-1}.
\end{equation}

A particular set of gauge invariant functions are the Wilson
loops.

If $C=(i_1, ...,i_p, i_1)$ is a loop in $R,$ with $(i_j,
i_{j+1})$ element of $L,$  we will define $U_C=\prod_{j=1}^p
U_{(i_j, i_{j+1})} .$

Let $\phi$ be any  central
function on $G,$  (i.e $\phi(xy)=\phi(yx), \forall x, y\in G$)
then $\phi(U_C)$ is a
gauge invariant function, which morever does not depend on the
departure point of $C.$

Let $\alpha$ be a representation of $G,$ then the character
$\chi_{\alpha}$ is a central map.

When $G=SU(n) , SO(n)$ it is custom to associate
to each plaquette, i.e to each elementary loop
$P=(i,j,k,l,i)$ a Boltzmann weight
\begin{equation}
w_{\beta}(U_P)=exp({\beta\over  n}Re (\chi_{f}(U_P)-n))\label{wbeta}
\end{equation}
where $f$ is the  fundamental representation of $G$    and $n$ the
dimension of
$f.$

Let $d\mu$ be the normalized Haar measure of $G$, this measure  is
right and left invariant.

We can define a  gauge invariant measure on the set of configurations
 by:

\begin{equation}
d\nu(U_{l\in L})=\prod_{P}w_{\beta}(U_P) \prod_{l\in L}d\mu (U_l),
\end{equation} where $P$ exhausts the set of all  plaquettes, $l$ the set
of links, and $U_l$ satisfy relation (\ref{uu=1}).

One is then interested in the evaluation of mean values such
as:\\
$<\psi( (U_l)_{l\in I}) d\nu>$ where $I$ is a finite set of links
and $\psi$ a function of the variables  $(U_l)_{l\in I}.$

It is also possible to define another Boltzmann weight \cite{Mi} which
includes all
the equivalency classes of  representations of $G$ and is  equivalent to
the weight (\ref{wbeta}) when
$\beta$ goes to infinity or in the continuum limit. This weight is defined
by:

\begin{equation}
w_{a,\beta}(U_P)=\sum_{\alpha \in Irr(G)} d_{\alpha} \chi_{\alpha}(U_P)
e^{-{
a  C_{\alpha}\over{\beta  n}}},\label{wabeta}
\end{equation}

where we have used the notation $Irr(G)$ for the set of equivalency classes
of
irreducible
representations of $G,$  $d_{\alpha}$ for the dimension of $\alpha,$ and
$C_{\alpha}$ for the value of the Casimir element in the representation
$\alpha.$

When $\beta$ goes to infinity the two weights approach  the $\delta$
function
 located at the unit element.

In two dimensions it is well known that Yang Mills theory is invariant
under area
preserving diffeomorphisms, and when the coupling constant of the
theory goes to zero then  this symmetry is enlarged to the group of all
diffeomorphisms
of the surface, i.e Yang Mills theory in two
dimensions is
 a topological field theory  \cite{Ru} \cite{Wi}.

The Boltzmann weight (\ref{wabeta}) is perfectly suited to describe  this
invariance
nature of
the theory because it satisfies an exact  block spin transformation, as
shown by
Migdal \cite{Mi} i.e :

\begin{equation}
\int w_{a,\beta}(xy) w_{a',\beta}(y^{-1}z) d\mu (y)=
w_{a+a',\beta}(xz). \end{equation}

Following the general philosophy of quantum geometry we will now redefine
algebraic
properties of
lattice gauge theory using only the algebra of functions on the set of
link variables and
on the gauge group.

Let us define for $z\in V,$ the group  $G_z=G\times \{z\}$ and $\hat
G=\prod_{z\in V} G_{z}.$

$\hat G$ is the gauge group.
Let us  now define for $\l \in L,$ the set $G_l=G\times \{l\}$ and ${\cal
L}=\prod_{l\in L}
G_{l},$ $\cal L$ is the set of link variables.  The gauge group $\hat G$
acts on the set ${\cal
L}$ by gauge transformations:

\begin{eqnarray}
\omega:&&\hskip 20pt {\cal L}\times \hat G \hbox to 95pt{\rightarrowfill}
{\cal
L}\nonumber\\
 & &(U_{(xy)})_{(xy)\in L} \times (g_{z})_{z\in V}  \mapsto
(g_x U_{(xy)}g_y^{-1})_{(xy)\in L}. \end{eqnarray}
The general principle of arrows reversing implies that if we define the
algebras
$\hat\Gamma={\cal F}(\hat G)=\bigotimes_{z\in V}{\cal F}(G_z)$ and
$\Lambda={\cal
F}({\cal L}),$ we have a coaction $\Omega$ of the Hopf algebra $\hat\Gamma$
on
$\Lambda$: \be
\Omega:\Lambda\rightarrow\Lambda\otimes \hat\Gamma.
\ee
It satisfies the usual axiom of right coaction:
\be
(\Omega\otimes id)\Omega=(id\otimes\Delta)\Omega\\
\ee
By construction it is a morphism of algebras:
\be
\Omega(AB)=\Omega(A)\Omega(B)\quad \forall\, A,B\in \Lambda
\ee

The algebra ${\cal F}(G_{(xy)})$  has a Hilbert basis which consists in the
matrix elements
$\alpha(U_{(xy)}){}^j_i$  where $\alpha$ is in $Irr(G),$ and the coaction
$\Omega$ on them can
be written:
\be
\Omega(\alpha(U_{(xy)}){}^i_j)= \alpha((g_x U_{(xy)}g_y^{-1})){}^i_j=
\sum_{p,q}\alpha(U_{(xy)}){}^p_q \alpha(g_x){}^i_p \alpha(g_y^{-1}){}^q_j
\ee
We will systematically use this point of view when dealing with the
q-deformed  version.

It is trivial to show that an element $\phi$ of $\Lambda$ is a gauge
invariant function if and only if:
\be
\Omega(\phi)=\phi\otimes 1.
\ee

In the next section we  study a q-analog of lattice gauge theory. The main
principle
which will guide us  throughout this section is   gauge covariance: in
technical terms,  we  define  a non commutative
deformation of the algebra $\Lambda={\cal F}({\cal L})$  such that a Hopf
algebra $\hat\Gamma$ (``the quantum gauge group'')
coacts on it.

\section{Gauge symmetry and algebra of gauge fields}

Let $\cal G$ be a  simple Lie algebra over $\bf C$ and let $q$ be a complex
number non root of unity
then the Hopf algebra  $A=U_q{\cal G}$ is quasitriangular. Let $Irr(A)$  be
the set of all
equivalency classes of  finite dimensional irreducible representations. In
each of these  classes
$\dot{\alpha}$   we will pick out a particular representative $\alpha.$
Let $V_{\alpha}$ be the
vector space on which acts the representation $\alpha.$

We will denote by $\bar{\alpha}$ (resp.  $\tilde{\alpha}$) the right (resp.
left ) contragredient representation associated to
$\alpha$ acting on $V_{\alpha}^{\star}$ and defined by:
$\bar{\alpha}={}^t\alpha \circ S $ (resp. $\tilde{\alpha}={}^t\alpha \circ
S^{-1}).$
We will also denote by $0$ the representation of dimension $1$ related to
the counit $\epsilon.$

Due to the  quasitriangularity of $A$ there
exists an invertible element $R\in A\otimes A$ (the universal R-matrix)
such that:
\bea
\Delta'(a)&=&R \Delta(a) R^{-1}\,\,\forall a\in A \\
(\Delta\otimes id)(R)&=&R_{13}R_{23}\\
( id\otimes\Delta)(R)&=&R_{13}R_{12}
\eea
Let us write $R=\sum_{i}a_i\otimes b_i,$ then the element
$u=\sum_{i}S(b_i)a_i$ is invertible
and satisfies \cite{D}:

\bea
&& S^2(x)=uxu^{-1}\label{u}\\
&& \mbox{$uS(u)$ is a central element}\nonumber\\
&&\sum_{i}b_i u a_i=\sum_{i}S(b_i)uS(a_i)=1.\label{bua}\\
&&\Delta(u)=(u\otimes u)(R_{21}R_{12})^{-1}=(R_{21}R_{12})^{-1}(u\otimes u)
\label{Deltau}
\eea

Moreover $U_{q}{\cal G}$ is a ribbon Hopf algebra \cite{RT}, which means
that it exists an
invertible element $v$ such that:
\bea
&&\mbox{$v$ is a central element}\nonumber\\
&& v^2=uS(u), \epsilon(v)=1, S(v)=v \label{v}\\
&&\Delta(uv^{-1})=uv^{-1}\otimes uv^{-1}
\eea

Let us define the group-like element $\mu=uv^{-1}$ and
$\mua=\alpha(\mu)\in  End(V_{\alpha})..$

 The $q-$dimension of $\alpha$ is defined by $[d_{\alpha}]=
 tr_{V_{\alpha}}(\mua)=tr_{V_{\alpha}}(\mua{}^{-1}).$
We also denote by $v_{\alpha}$ the complex number $\alpha(v),$
and $\sua=\alpha(u).$

We  define $\Rab=(\alpha\otimes \beta)(R)\in \End(V_{\alpha}\otimes
V_{\beta}).$  Let $(\ea{}_i)_{i=1...dim V_{\alpha}}$ be a particular basis
of $V_{\alpha},$  and
 $(\ea{}^i)_{i=1...dim V_{\alpha}}$ its dual basis.
We will define the linear forms $\ga{}^j_i=<\ea{}^j\vert \alpha(.)\vert
\ea{}_i>.$

The existence of $R$ implies that they satisfy the exchange relations:

\begin{equation}
\sum_{i,k}\Rab{}^{pq}_{ik} \ga{}^i_j \gb{}^k_l =
\sum_{m,n}\gb{}^q_n \ga{}^p_m \Rab{}^{mn}_{jl}
\end{equation}
which can be simply written:
\begin{equation}
\Rab_{12} \ga_1 \gb_2 = \gb_2 \ga_1 \Rab_{12},\label{Rgg}
\end{equation}
using the convenient notation: $\ga=\sum_{i,j} \ea_i\otimes \ea{}^j\otimes
\ga{}^i_j.$
This relation is also equivalent to:
\begin{equation}
\Rab{}^{-1}_{21} \ga_1 \gb_2 = \gb_2 \ga_1 \Rab{}^{-1}_{21}.\label{Rgg'}
\end{equation}

Let $\Gamma$ be the restricted dual of $U_q({\cal G}),$ it is by
construction
a Hopf algebra, generated as a vector space by the elements $\ga{}^i_j.$

The action of the coproduct on these elements is:

\begin{equation}
\Delta (\ga{}^i_j)=
\sum_k <\ea{}^i\vert \alpha(.)\vert \ea{}_k ><\ea{}^k\vert \alpha(.)\vert
\ea{}_j >= \sum_k \ga{}^i_k \otimes \ga{}^k_j\label{coproduit}.
\end{equation}

$V_{\alpha}$ can be endowed with a structure of right  comodule over
$\Gamma:$
\be
\Delta_{\alpha}(\ea_i)=\sum_{j} \ea_j\otimes \ga{}^j_i.
\ee

 We also have
\be
\Delta_{\bar\alpha}(\ea{}^i)=\sum_{j} \ea{}^j\otimes  S(\ga{}^i_j)=
\sum_{j} \ea{}^j\otimes  \gabar{}^j_i.
\ee

Let $\alpha, \beta$ be two fixed elements of $Irr(A),$  because $q$ is not
a
root of unity, finite dimensional representations are completely
reducible:
\be
\alpha\otimes\beta= \bigoplus_{\gamma\in Irr(A)}N_{\alpha\beta}^{\gamma}\,
\gamma
\ee

Let us define, for each $\gamma,$
$(\psi^{\alpha,\beta}_{\gamma,m})_{m=1...N_{\alpha\beta}^{\gamma}}$  a
basis of $Hom_A(V_{\alpha} \otimes
V_{\beta},V_{\gamma})$
 and  $(\phi^{\gamma,m}_{\alpha,\beta})_{m=1...N_{\alpha\beta}^{\gamma}}$ a
basis of $Hom_A(V_{\gamma},V_{\alpha} \otimes
V_{\beta}):$

\begin{equation}
V_{\alpha} \otimes V_{\beta} \buildrel {\psi^{\alpha,\beta}_{\gamma,m}}
\over \rightarrow
V_{\gamma} \buildrel {\phi^{\gamma,m'}_{\alpha,\beta}} \over \rightarrow
V_{\alpha} \otimes V_{\beta}
\end{equation}

We can always choose them such that they own the properties:

\begin{eqnarray}
\sum_{m,\gamma}\phi^{\gamma,m}_{\alpha,\beta}
\psi^{\alpha,\beta}_{\gamma,m}&=&
id_{V_{\alpha} \otimes V_{\beta}}\\
 \psi^{\alpha,\beta}_{\gamma,m}\phi^{\gamma',m'}_{\alpha,\beta} &=&
\delta^m_{m'}
 \delta^{\gamma'}_{\gamma'}id_{V_{\gamma}}.
\end{eqnarray}

It is easy to check that we can always choose:
\begin{eqnarray}
\phi^{\gamma,m}_{\beta,\alpha} &=& \lambda_{\alpha\beta\gamma}
P_{12}\Rab_{21}^{-1}\phi^{\gamma,m}_{\alpha,\beta} \label{cg3}\\
\psi_{\gamma,m}^{\beta,\alpha} &=& \lambda_{\alpha\beta\gamma}^{-1}
\psi_{\gamma,m}^{\alpha,\beta} \Rab_{21}  P_{12}\label{cg4}
\end{eqnarray}
where
$\lambda_{\alpha\beta\gamma}=(v_{\alpha}v_{\beta}v_{\gamma}^{-1})^{1/2}.$

We can define the Clebsch-Gordan coefficients by

\begin{equation}
\psi^{\alpha,\beta}_{\gamma,m}(\ea_a \otimes \eb_b) =
\sum_{c} \ec_{c} \CGpsi{\alpha}{a}{\beta}{b}{\gamma}{c}{m}
\end{equation}

\begin{equation}
\phi^{\gamma,m}_{\alpha,\beta}(\ec_c) =
 \sum_{a,b} \ea_a \otimes \eb_b \CGphi{\alpha}{a}{\beta}{b}{\gamma}{c}{m}.
\end{equation}

 We then obtain the relation:
\begin{equation}
\ga{}^i_j \gb{}^k_l = \sum_{\gamma,m,p,q}
\CGphi{\alpha}{i}{\beta}{k}{\gamma}{p}{m}
 \gc{}^p_q
\CGpsi{\alpha}{j}{\beta}{l}{\gamma}{q}{m}.
\end{equation}

This relation can also be written:
\be
\ga_1\gb_2=\sum_{\gamma,m}\phi^{\gamma,m}_{\alpha,\beta} \gc
\psi^{\alpha,\beta}_{\gamma,m}.
\ee

Let us now specialize $\cal G$ to be a Lie algebra of type $A_n, B_n, C_n,
D_n$  and $f$ be a
fundamental representation of $U_{q}(\cal G)..$ In this case the restricted
dual is the
Hopf algebra generated by  $\gf$ satisfying the relation:

\be
\Rff_{12}\, \gf_1 \gf_2 =\gf_2 \gf_1\,\Rff_{12}  \,{\rm and}\,\,\,
\nonumber\left\{\begin{array}{ll}
 {\rm det}_{q}(\gf)=1\ & \mbox{if }\, {\cal G}=A_n,\nonumber\\
\gf C {}^t\gf C^{-1}=C{}^t\gf C^{-1}\gf=I & \mbox{if }\,{\cal G}=B_n, C_n,
D_n
\nonumber\end{array}\right.
\nonumber
\ee

where ${\rm det}_{q}$ is the quantum determinant and $C$ is the quantum
quadratic form defined, for example, in \cite{FRT}.

The antipode is defined to be the  antimorphism of algebra satisfying:
\be
\sum_{j}\gf{}^i_j S(\gf{}^j_k)=\sum_{j}S(\gf{}^i_j) \gf{}^j_k=\delta_k^i.
\ee

An explicit description of $S$ is obtained using the quantum analog of
Cramers formulas
 \cite{FRT}.
The matrix elements $\ga{}^j_i$ are expressed as polynomials in the matrix
elements
$\gf{}^j_i,$ these polynomials can be computed using the Clebsch-Gordan
coefficients.

Because of the relation (\ref{coproduit}) they satisfy:
\be
\sum_{j}\ga{}^i_j S(\ga{}^j_k)=\sum_{j}S(\ga{}^i_j) \ga{}^j_k=\delta_k^i.
\ee

We will now define a non commutative analog of the  gauge symmetry Hopf
algebra
$\hat \Gamma$ and  a non commutative analog of the gauge field algebra
$\Lambda.$
\medskip

Let $\Sigma$ be a compact connected triangulated oriented  Riemann surface
with
boundary $\partial\Sigma$ and let  $(F_i)_{i=1,...,n_F}$ be the
oriented faces of  $\Sigma.$ Let us denote by $L$ the set of edges counted
with the orientation induced by the orientation of the corresponding faces.
We have $L=L^i\cup L^b,$ where $L^i, L^b$ are respectively the set of
interior
edges and boundary edges $(\partial\Sigma=L^b)$.

Finally let us also define $V$ to be the set of points (vertices)
of
this triangulation, $V=V^i\cup V^b$, where $V^i, V^b$ are respectively the
set
of interior vertices and boundary vertices.

If $l$ is an oriented link it will be convenient to write $l=(x,y)$ where
$x$ is the departure point
of $l$ and $y$ the end point of $l.$ We will write $x=d(l)$ and $y=e(l).$

 It is important to notice that
the definition of a triangulation imposes  $x\not=y.$ This property implies
that for any link $l$
incident to a vertex $z$ it is possible to determine unambiguously   if $z$
is the departure or  end
point of $l.$ We will moreover assume that a link $l$ is completely
characterized by its departure and
end points.

 If l is an interior edge then we  define $\bar l$ to be the
corresponding edge with opposite orientation.

\bd[gauge symmetry algebra]
Let us define for $z\in V,$ the Hopf algebra  $\Gamma_z=\Gamma\times \{z\}$
and $\hat
\Gamma=\bigotimes_{z\in V} \Gamma_{z}.$ This Hopf algebra  will be called
the gauge
symmetry algebra.
\ed

In order to define the non commutative algebra of gauge fields we will have
to
endow the triangulation with an additional structure, an  order on the set
of links incident to each vertex, the {\sl cilium order}.
This can be achieved using the formalism of ciliated fat graph described by
V.V.Fock and A.A.Rosly \cite{FR}.

\bd[Ciliation]
A ciliation of the triangulation is an assignment of a cilium
$c_z$ to each vertex $z$   which consists in a non zero tangent vector at
z.
The orientation of the Riemann surface
defines a canonical cyclic order of the links admitting $z$ as departure or
end
point. Let $l_1, l_2$ be links incident to a common vertex $z,$
the  partial cilium order $< $ is defined by:

 $l_1<l_2$ if $l_1\not=l_2, {\bar l}_2$ and  the unoriented vectors
 $c_z,l_1,l_2$ appear
in the cyclic order defined
by the orientation of the surface.
\ed

We will assume in the rest of this work that the triangulated Riemann
surface $\Sigma$ is endowed with a  ciliation.

\bd[Gauge fields algebra]
We shall now define the algebra of gauge fields $\Lambda$ to be the algebra
generated by the formal variables $\ua(l)^i_j$ with $l\in L,\alpha\in
Irr(A),i,j=1\cdots dim V_{\alpha}$ and satisfying the following relations:

\smallskip

{\bf Commutation rules}
\begin{eqnarray}
& &\ua (x,y)_1  \ub (z,y)_2\Rab_{12} =
 \ub (z,y)_2 \ua (x,y)_1 \label{EE}\\
 & &\ua (x,y)_1 {}\Rab_{12}^{-1} \ub (y,z)_2 =\ub (y,z)_2 \ua
(x,y)_1\label{ES}\\
& &\Rab_{12} \ua (y,x)_1  \ub (y,z)_2 =  \ub (y,z)_2 \ua
(y,x)_1\label{SS}\\
& &\, \forall \,\,(y,x), (y,z) \in L\, x\not= z \,\,\,{\rm and}\,\,\,
(x,y)<(y,z)\nonumber
\eea
\bea
& &\Rab_{12} \ua (x,y)_1  \ub (x,y)_2 \Rab{}^{-1}_{21}=
 \ub (x,y)_2 \ua (x,y)_1 \label{EEegaux}\\
& &\forall \,\,(x,y)\in L, \nonumber\\
& &\ua(x,y)\ua(y,x)=1\label{ES=1}\\
& &\forall \,\,(x,y)\in L^i, \nonumber
\eea
\bea
& &\ua (x,y)_1 \ub (z,t)_2 = \ub (z,t)_2  \ua (x,y)_1 \label{Udisjoint}\\
& &\forall\,\, x, y, z, t \mbox{   pairwise distinct in}\, V\nonumber
\end{eqnarray}

\smallskip

{\bf Decomposition rules}
\bea
\ua(x,y)_1 \ub(x,y)_2&=&\sum_{\gamma,m}\phi^{\gamma,m}_{\alpha,\beta}
\uc(x,y)\psi^{\beta,\alpha}_{\gamma,m}P_{12}, \label{UCG}\\
\uo(x,y)&=&1,\,\,\forall (x,y)\in L
\eea

where we have systematically used the notation:
\be
\ua (x,y)=\sum_{i,j} \ea_i\otimes \ea{}^j\otimes \ua (x,y){}^i_j.
\ee

\vskip 15pt
\ed

This algebra was recently introduced by A.Yu.Alekseev, H.Grosse and
V.Scho\-merus
in \cite{AGS}.

The defining relations of the gauge fields algebra are obtained  by
demanding
a property of covariance under the coaction of
the gauge algebra $\hat\Gamma$ defined as follows:

\medskip

\bd[Gauge covariance]
$\Lambda$ can be endowed with a right comodule structure
$\Omega:\Lambda\rightarrow \Lambda\otimes \hat\Gamma$ such that
\begin{enumerate}
\item $\Omega$ is a  morphism of algebras.
\item The action of $\Omega$ on the elements $\ua(x,y){}^i_j$ is defined by
the formula:
\end{enumerate}
\be
\Omega(\ua(x,y))=(\Delta_{\alpha}\otimes \Delta_{\bar\alpha}\otimes
id)(\ua(x,y)).
\ee

This expression can be expressed in terms of components as:
\be
\Omega(\ua(x,y){}^i_j)=\sum_{p, q}\ua(x,y){}^p_q\otimes (\ga{}^{x})^{i}_p
S((\ga{}^{y})^{q}_j),
\ee

where $(\ga{}^{x})^{i}_p$ is the image of the element $\ga{}^{i}_p$ by the
canonical injections
$\Gamma\hookrightarrow \Gamma_{x}\hookrightarrow \hat\Gamma.$
\ed

{\sl Proof:}

It is straightforward to show that the relations of definition of $\Lambda$
are compatible
with the definition of $\Omega.$   We shall verify in detail that
relations (\ref{EE}, \ref{UCG})
are indeed covariant under the coaction of $\hat \Gamma.$ Covariance of
other
relations can be checked using the same scheme.

\smallskip
Verification of the covariance of relation (\ref{EE}):
\bea
\Omega(\ua (x,y)_1 \ub (z,y)_2 \Rab_{12})&=& \Omega(\ua (x,y)_1)
\Omega (\ub (z,y)_2) \Rab_{12}\nonumber\\
&=&\ga{}^x_1 \ua(x,y)_1 S(\ga{}^y)_1 \gb{}^z_2 \ub(z,y)_2
S(\gb{}^y)_2\Rab_{12}\nonumber\\
&=&\ga{}^x_1 \gb{}^z_2 \ua(x,y)_1\ub(z,y)_2 S(\ga{}^y)_1
S(\gb{}^y)_2\Rab_{12}\nonumber\\
&=&\ga{}^x_1 \gb{}^z_2 \ua(x,y)_1\ub(z,y)_2 \Rab_{12} S(\gb{}^y)_2
S(\ga{}^y)_1 \nonumber\\
&=&\ga{}^x_1 \gb{}^z_2 \ub(z,y)_2 \ua(x,y)_1 S(\gb{}^y)_2 S(\ga{}^y)_1
\nonumber\\
&=&\gb{}^z_2 \ub(z,y)_2 S(\gb{}^y)_2 \ga{}^x_1\ua(x,y)_1  S(\ga{}^y)_1
\nonumber\\
&=&\Omega( \ub (z,y)_2 \ua (x,y)_1 ).\nonumber
\eea

Verification of  covariance of relation (\ref{UCG}):
\bea
\Omega(\ua (x,y)_1 \ub (x,y)_2)&=&\Omega(\ua (x,y)_1)\Omega( \ub
(x,y)_2)\nonumber\\
&=&\ga{}^x_1 \ua(x,y)_1 S(\ga{}^y)_1\gb{}^x_2 \ub(x,y)_2 S(\gb{}^y)_2
\nonumber\\
&=&\ga{}^x_1\gb{}^x_2 \ua(x,y)_1 \ub(x,y)_2  S(\ga{}^y)_1 S(\gb{}^y)_2
\nonumber\\
&=&\ga{}^x_1\gb{}^x_2 \ua(x,y)_1 \ub(x,y)_2 \Rab_{21}{}^{-1}
S(\ga{}^y_1\gb{}^y_2 )
\Rab_{21}\nonumber
\eea
\bea
&=&\sum_{\gamma,\gamma',\gamma''\atop m,m',m''}
\phi^{\gamma,m}_{\alpha,\beta} \gc \psi^{\alpha,\beta}_{\gamma,m}
\phi^{\gamma',m'}_{\alpha,\beta} \uc(x,y) \psi^{\beta,\alpha}_{\gamma',m'}
P_{12}\Rab_{21}{}^{-1}
\phi^{\gamma'',m''}_{\alpha,\beta} S(\gc)
\psi^{\alpha,\beta}_{\gamma'',m''}
\Rab_{21}\nonumber\\
&=&\sum_{\gamma,\gamma',\gamma''\atop m,m',m''}
\phi^{\gamma,m}_{\alpha,\beta} \gc \psi^{\alpha,\beta}_{\gamma,m}
\phi^{\gamma',m'}_{\alpha,\beta} \uc(x,y) \psi^{\alpha,\beta}_{\gamma',m'}
\lambda_{\alpha\beta\gamma'}^{-1}
\phi^{\gamma'',m''}_{\alpha,\beta} S(\gc)
\psi^{\alpha,\beta}_{\gamma'',m''}
\Rab_{21}\nonumber\\
&=&\sum_{m,\gamma}\phi^{\gamma,m}_{\alpha,\beta}
\Omega(\uc(x,y)) \psi^{\alpha,\beta}_{\gamma,m} \Rab_{21}
\lambda_{\alpha\beta\gamma}^{-1}\nonumber\\
&=&\Omega(\sum_{m,\gamma}\phi^{\gamma,m}_{\alpha,\beta}
\uc(x,y) \psi^{\beta,\alpha}_{\gamma,m}P_{12}).\nonumber
\eea
$\Box$

 At this point we have many comments to make.
\smallskip

Remark 1.
We just have seen that the braiding relation (\ref{Rgg}) is compatible with
 the
exchange relation (\ref{EE}).
But relation (\ref{Rgg}) can also be written (\ref{Rgg'}), this
would as well implies an exchange relation of the type (\ref{EE}) with
$R_{12}$ exchanged with $R^{-1}_{21}.$
The aim of the ciliation is precisely to solve this ambiguity. Indeed if
$i=(x,y)$ and
$j=(z,y)$ with $i<j$ then the relation
\be
\ua (i)_1  \ub (j)_2\Rab_{12} =
 \ub (j)_2 \ua (i)_1
 \ee
 implies $\ua (i)_1  \ub (j)_2\Rab{}^{-1}_{21} =
 \ub (j)_2 \ua (i)_1$ for $i>j.$ This can be easily checked by applying the

 permutation operator.

\smallskip

Remark 2.
The commutation relations given in the definition of $\Lambda$ are not
minimal.
The exchange relations of the type (\ref{EE}, \ref{ES}, \ref{SS}) are
related to each other
using relation (\ref{ES=1}) when links are interior links.
If $(x,y)$ is an interior link then we also have from relation
(\ref{EEegaux}):

 \bea
& &\ua (x,y)_1\Rab_{21} \ub (y,x)_2 = \ub (y,x)_2 \Rab_{12} \ua (x,y)_1
\label{ESopp1}\\
& &\ua (x,y)_1\Rab_{12}^{-1} \ub (y,x)_2 = \ub (y,x)_2 \Rab_{21}^{-1} \ua
(x,y)_1 \label{ESopp2}
\eea
It is also easy to see that (\ref{UCG}) implies relations (\ref{EEegaux}).
Moreover we
have a relation between $\ua(y,x)$ and $\uabar(x,y)$ which can be shown to
hold
using (\ref{UCG},\ref{ES=1}):

\bp
If $(x,y)$ is an interior link, we have
\be
\ua(y,x)=\mua{}^{-1}{}^t\uabar(x,y) \label{ul=ubarl'1}
\ee
(This is the relation (4.8) of \cite{AGS}.)
\ep

{\sl Proof:}

 We  first have to describe in details the map
$\psi^{\bar{\alpha},\alpha}_{0}$ and
$\phi^{0}_{\bar{\alpha},\alpha}$ where as usual $0$ denote the trivial
representation of
dimension $1.$
$\psi^{\bar{\alpha},\alpha}_{0}$ is  the usual canonical map:
\begin{eqnarray}
\psi^{\bar{\alpha},\alpha}_{0}:&&V_{\bar\alpha}\otimes
V_{\alpha}\rightarrow  {\bf C}\nonumber\\
&&\xi\otimes x\mapsto <\xi,x>.
\end{eqnarray}
$\phi^{0}_{\bar{\alpha},\alpha}$ is the  map:
\begin{eqnarray}
\phi^{0}_{\bar{\alpha},\alpha}:&&{\bf C}\rightarrow   V_{\bar\alpha}\otimes
V_{\alpha}\nonumber\\
&&\lambda\mapsto \lambda{1\over tr(u_{\alpha}^{-1})}\sum_{i}
\ea{}^{i}\otimes u_{\alpha}^{-1}\ea_i.
\end{eqnarray}
This is an easy consequence of the identity $S(x)=u S^{-1}(x) u^{-1}.$ The
normalization of
$\phi^{0}_{\bar{\alpha},\alpha}$ is chosen
such that
$\psi^{\bar{\alpha},\alpha}_{0}\phi^{0}_{\bar{\alpha},\alpha}=id_{\bf C}.$

Let $(x,y)\in L,$
the decomposition rule (\ref{UCG}) implies that
\be
\psi^{\bar{\alpha},\alpha}_{0}\uabar(x,y)_1\ua(x,y)_2=
\psi^{\alpha,\bar{\alpha}}_{0}P_{12}=
\psi^{\bar{\alpha},\alpha}_{0}\Rabara_{21}v_{\alpha}^{-1},
\ee
in terms of components this can be written:
\be
\sum_i \uabar (x,y){}^{i}_j \ua(x,y){}^i_l=\mua{}^j_l.
\ee
This relation can also be written
\be
\ua(y,x){}^l_i=\sum_{j}\uabar (x,y)^{i}_j (\mua{}^{-1})^l_j,
\ee
which is exactly relation (\ref{ul=ubarl'1}).
$\Box$

\medskip

Remark 3.
The associativity ``constraints''  coming from exchange relations between
elements of the algebra  are automatically satisfied from
quasitriangularity
of $U_{q}(\cal{G}).$

\smallskip

Remark 4.
In the classical case the variables $U_{ij}$ are elements of the group $G,$
in this case
the gauge fields variables are of the same type as elements of the gauge
group. In the q-deformed
case this doesnot hold true as can be seen from the commutation relations
in $\Lambda.$
Let $X\subset L$ and $\Lambda_X$ the subalgebra of $\Lambda$ generated by
$\ua(l)$ with
$l\in X$ and $\alpha\in Irr(A).$ The q-deformed lattice theory is a
nonultralocal theory in
the sense that the algebras $\Lambda_X$ and $\Lambda_Y$  are pointwise
commuting if
$X$ and $Y$ have no link incident to a same vertex.

\medskip

Let $M=\{l_1,\cdots, l_p\}$ be a subset of links such that $M$ contains the

 boundary links and each of the interior link in one and only one
orientation.

It is trivial to show using the commutation  relations in  the algebra
$\Lambda$ and the relations (\ref{ul=ubarl'1},\ref{ul=ubarl'2}) that
$\{\prod_{j=1}^{p}\Ua{j}(l_{j}){}^{m_j}_{n_j}\}$,
$\alpha_j\in Irr(A), m_j,n_j=1\cdots d_j$ is a generating family
of the vector space
$\Lambda.$

We will assume in the rest of the paper that this family is a basis of
$\Lambda.$ This assumption is quite natural and perhaps can it be proved
using a
representation of the algebra $\Lambda$  or using techniques like the
Diamond Lemma.
What can easily be shown is  that this property is independant of the
choice of  $M.$ We will use the assumption of independance of this family
of vectors
in order to build an
analogue of the Haar measure in section (5).

\section{Wilson Loops}
In this section we  define elements of the algebra $\Lambda,$ called ``{\sl
Wilson loops}''
attached to loops on the lattice and owning the following properties: gauge
invariance and
invariance under departure point (we will call this last property {\sl
cyclicity}).
\medskip

A loop $C$ of length $k=l(C)$ is said {\sl simple} if $C=(x_1,\cdots
x_k,x_{k+1}=x_{1})$
with
$x_1,\cdots, x_k$ pairwise distincts. From now we allow us to identify
$x_{n+k}$
 and $x_{n}$ for all $n$.

Let us define  the sign $\epsilon(x_i, C)$ to be $1$ (resp. $-1$) if
$(x_{i-1},x_i)<(x_i,x_{i+1})$ (resp. $(x_{i-1},x_i)>(x_i,x_{i+1})$) and
denote
$N(C)$  the cardinal of the set

\noindent$\{i \in \{1..k\}/ (x_{i-1},x_i)<(x_i,x_{i+1})\}$.

If C is a contractile loop, $N(C)$  is simply the number of
cilia located at the vertices of $\{x_1,\cdots, x_k\}$
 and directed inside the domain
enclosed by the loop $C.$

We have the relation:
$2N(C)=l(C)+\sum_{x\in C}\epsilon(x,C).$

Let $C_1$ and $C_2$ be two simple loops of length greater than three with
common edges such that
$C_1$ and $C_2$ have opposite orientation on these edges. We will denote
$C_1\#C_2$ to be the loop obtained by gluing $C_1$ and $C_2$ along their
common part and removing these common edges.

\bd
Let ${\cal C}_k$   be the set of loops of length $k$ $(k\geq 3)$ i.e the
set of

$(k+1)-$~uplets
 $C=(x_1,\cdots,x_k,x_{k+1}=x_1)$
 where $(x_{i}, x_{i+1})\in L$).
The set of loops is defined by ${\cal C}=\bigcup_{k\geq 2}{\cal C}_k.$
We also define ${\cal C}^s$ to be the set of simple loops.

Let us define for arbitrary adjacent
links $(x,y), (y,z)$ the matrix:

\be
{\cal R}^{\alpha\beta}(x,y,z)=\left\{\begin{array}{ll}
 \Rab{}^{-1}_{12} & \mbox{if $(x,y)<(y,z)$}\\
\Rab{}_{21} & \mbox{if $(x,y)>(y,z)$}
\end{array}\right.
\ee

We define the Wilson loop
in the representation $\alpha$ attached to $C\in{\cal C}^s$ by:
\be
W^{\alpha}(C)=
\omega_{\alpha}(C) v_{\alpha}tr_{1..k}(\mua{}^{\otimes k}\sigma^{(k)}
(\prod_{j=1}^{k-1}\ua(x_j, x_{j+1})_j{\cal R}_{jj+1}^{\alpha\alpha}(x_j,
x_{j+1}, x_{j+2}))
 \ua(x_k,x_1)_{k}).\nonumber \ee

where the expression $tr_{1..k}$ means the trace over the space
$V_{\alpha}^{\otimes k},$ $\sigma^{(k)}$ is the permutation operator
$\sigma^{(k)}=P_{kk-1}\cdots P_{21},$ and $\omega_{\alpha}(C)$ is a non
zero complex number
depending only on the  distribution of cilia along the loop.
\ed

\medskip
\bp
 Wilson loops $W^{\alpha}(x_1,\cdots,x_k)$ are gauge invariant and
cyclic invariant i.e  $W^{\alpha}(x_1, x_2,\cdots,x_k)=W^{\alpha}(x_2,
\cdots, x_k, x_1).$
We have the following important relation:
\be
 W^{\alpha}(C)=\rho_{\alpha}(x_1,C) \omega_{\alpha}(C)
 tr_{V_{\alpha}} (\mua \ua(x_{1},x_{2})
\cdots \ua(x_{k},x_{1}))
 \ee

where
$\rho_{\alpha}(x_1,C)=v_{\alpha}^{1+\sum_{x\not=x_1}\epsilon(x_1,C)}.$

\ep
Proof:

Let us show first the cyclicity property.
 From the definition of $\cal{R},$ the commutation rules (\ref{ES})
can be written:
\be
\ua(x_{j-1}, x_{j})_1{\cal R}_{12}^{\alpha\alpha}(x_{j-1}, x_{j}, x_{j+1}))
\ua(x_j,x_{j+1})_{2}=\ua(x_j,x_{j+1})_2 \ua(x_{j-1}, x_j)_1\nonumber.
\ee
We first have:
\bea
\noindent &&\prod_{j=1}^{k-1}(\ua(x_j, x_{j+1})_j{\cal
R}_{jj+1}^{\alpha\alpha}(x_j, x_{j+1}, x_{j+2}))
 \ua(x_k,x_1)_{k}=\nonumber\\
&=&\ua(x_1, x_2)_1{\cal R}_{12}^{\alpha\alpha}(x_1, x_2,
x_3)\ua(x_2,x_3)_{2}
\cdots{\cal R}_{k-1k}^{\alpha\alpha}(x_{k-1}, x_k,
x_{k+1})\ua(x_k,x_1)_{k}\nonumber\\
&=&\ua(x_2,x_3)_{2}\ua(x_1, x_2)_1{\cal R}_{23}^{\alpha\alpha}(x_{2},
x_3,x_{4})
\cdots{\cal R}_{k-1k}^{\alpha\alpha}(x_{k-1},
x_k,x_{k+1})\ua(x_k,x_1)_{k}\nonumber\\
&=&\ua(x_2,x_3)_{2}{\cal R}_{23}^{\alpha\alpha}(x_{2}, x_3,x_{4})
\cdots{\cal R}_{k-1k}^{\alpha\alpha}(x_{k-1}, x_k,x_{k+1})\ua(x_1,
x_2)_1\ua(x_k,x_1)_{k}\nonumber\\
&=&\ua(x_2,x_3)_{2}\cdots{\cal R}_{k-1k}^{\alpha\alpha}(x_{k-1},
x_k,x_{1})\ua(x_k,x_1)_{k}
{\cal R}_{k1}^{\alpha\alpha}(x_{k}, x_1,x_2)\ua(x_1, x_2)_1\nonumber\\
&=&\prod_{j=2}^{k}(\ua(x_j, x_{j+1})_j{\cal R}_{jj+1}^{\alpha\alpha}(x_j,
x_{j+1}, x_{j+2}))
 \ua(x_1,x_2)_{1}\nonumber.
\eea
Using the identity
\be
P_{kk-1}\cdots P_{32}P_{21} = P_{1k}P_{kk-1}\cdots P_{32},
\ee
we finally obtain:
\be
W^{\alpha}(x_1,x_2\cdots,x_k)=W^{\alpha}(x_2,\cdots,x_k,x_1),
\ee
which ends the proof of  cyclicity.

\smallskip
Using the following formula:
\be
tr_{1 \cdots k}(P_{kk-1}\cdots P_{32}P_{21} A^{(1)}_{1} \cdots A^{(k)}_{k})
=
 tr(A^{(1)} \cdots
A^{(k)}),
 \ee
where $A^{(1)}, \cdots, A^{(k)}$ are elements of $End(V_{\alpha})$
and the usual properties on $u$ recalled in (\ref{bua},\ref{v}), we can
easily  write
$W^{\alpha}(x_1, x_2,\cdots, x_k)$ in the more convenient form:
\be
 W^{\alpha}(C)= v_{\alpha}^{1+\sum_{x\not=x_1}\epsilon(x,C)}
\omega_{\alpha}(C) tr_{V_{\alpha}} (\sua \ua(x_{1},x_{2})
\cdots \ua(x_{k},x_{1})).
 \ee

 Gauge invariance is more explicit in this form. Indeed we have:
\bea
&&\Omega(v_{\alpha}^{-1-\sum_{x\not=x_1}\epsilon(x,C)}
\omega_{\alpha}(C)^{-1} W^{\alpha}(x_1,x_2\cdots,x_k)) = \\
&=&\Omega(\sua{}^{i_{k+1}}_{i_1}
\prod_{j=1}^{k}\ua(x_{j},x_{j+1})^{i_j}_{i_{j+1}})\nonumber\\
& &\mbox{(we have used an implicit sum over repeated indices)}\nonumber\\
&=&  \sua{}^{i_{k+1}}_{i_1}
\prod_{j=1}^{k}(\ga{}^{x_j})^{i_j}_{p_j}\ua(x_{j},x_{j+1})^{p_j}_{q_{j+1}}
S(\ga{}^{x_{j+1}})^{q_{j+1}}_{i_{j+1}}\nonumber\\
&=&
(\ga{}^{x_1})^{i_1}_{p_1}
S(\ga{}^{x_{1}})^{q_{k+1}}_{i_{k+1}}\sua{}^{i_{k+1}}_{i_1}
\prod_{j=2}^{k}S(\ga{}^{x_{j}})^{q_{j}}_{i_{j}}(\ga{}^{x_j})^{i_j}_{p_j}
\prod_{j=1}^{k} \ua(x_{j},x_{j+1})^{p_j}_{q_{j+1}}
\nonumber\\
&=&
(\ga{}^{x_1})^{i_1}_{p_1}\sua{}^{q_{k+1}}_{i_{k+1}}
S^{-1}(\ga{}^{x_{1}})^{i_{k+1}}_{i_{1}}
\prod_{j=2}^{k} \delta_{p_j}^{q_j} \prod_{j=1}^{k}
\ua(x_{j},x_{j+1})^{p_j}_{q_{j+1}} \nonumber\\
&=& \sua{}^{q_{k+1}}_{i_{k+1}} \delta^{i_{k+1}}_{p_1}
\prod_{j=2}^{k} \delta_{p_j}^{q_j} \prod_{j=1}^{k}
\ua(x_{j},x_{j+1})^{p_j}_{q_{j+1}} \nonumber\\
&=& v_{\alpha}^{-1-\sum_{x\not=x_1}\epsilon(x,C)} \omega_{\alpha}(C)^{-1}
W^{\alpha}(x_1,x_2\cdots,x_k)\otimes 1.\nonumber
\eea
which shows that $W^{\alpha}(x_1,x_2\cdots,x_k)$ is a gauge invariant
element

$\Box$

\medskip

Remark. We should make here a comment on the notion of trace and q-trace
following
the remark of Drinfeld in \cite{D}. Let $\ga$ be the matrix of elements
of $\Gamma.$ We can define $tr(\ga)=\sum_{i}\ga{}^i_i$ (the ordinary trace)

and $tr_{q}(\ga)=tr(\sua \ga)$ (the q-trace). It is a well known result
that the trace is cyclic but not Ad-invariant and on the contrary the
q-trace
is Ad-invariant but not cyclic. We need both of these properties to build
a  consistent q-Yang Mills type theory. One of the great benefit of the
algebra
$\Lambda$ is that the q-trace defined on it, is both cyclic and
Ad-invariant.

\medskip

A {\sl path} $P$ of length $k$ is a $(k+1)-$~uplet
 $P=(x_1,\cdots,x_{k+1})$
 where $(x_{i}, x_{i+1})\in L.$
In the proof of the next proposition we will have to extend the definition
of
 Wilson loop to a path $P$  by:
\be
M^{\alpha}(P)=
\prod_{j=1}^{k}\ua(x_j, x_{j+1})
\ee
When $P$ is a simple closed loop we of course have
$W^{\alpha}(P)=tr_{V^{\alpha}}(u_{\alpha}M^{\alpha}(P)).$
\medskip

We will now study the commutation relations of the Wilson loops.

\smallskip

Let us consider two loops $C$ and $C'$.
Let $C\cap C'$ be the set of common vertices of $C$ and $C'.$
$C\cap C'$ is naturally a disconnected union of paths
 $P^{(1)},.., P^{(k)}$ (resp  $P'{}^{(1)},.., P'{}^{(k)}$ ) with the
orientation induced by
 $C$ (resp. $C').$ For each of these paths we define
$L^{(1)},.., L^{(k)}$ (resp  $L'{}^{(1)},.., L'{}^{(k)}$ ) to be the paths
obtained from
 $P^{(1)},.., P^{(k)}$ (resp  $P'{}^{(1)},.., P'{}^{(k)}$ )
 by adding the neighbour vertices of the latter in $C$ (resp  $C'$).
Let us now consider a connected component $P^{(i)}$ of $C\cap C'$ and
its related paths
$L^{(i)}$ and $L'{}^{(i)}.$
$L^{(i)}$ (resp.$L'{}^{(i)}$) is a set of links
$(l_{0}^{(i)},l_{1}^{(i)},..,l_{n_i}^{(i)},l_{n_i+1}^{(i)})$
(resp.$(l_{0}'{}^{(i)},l_{1}^{(i)},..,l_{n_i}^{(i)},l'{}^{(i)}_{n_i+1}).$
$L^{(i)}$ is said to be a crossing zone if one of the following condition
is fullfilled:
\begin{enumerate}
\item \mbox{$n_i\not=0$, $l_{0}^{(i)}<l_{0}'{}^{(i)}<l_{1}^{(i)}$
and $l_{n_i}^{(i)}<l_{n_i+1}^{(i)}<l'{}^{(i)}_{n_i+1}$ or cyclic.perm.}
\item \mbox{$n_i\not=0$, $l_{0}'{}^{(i)}<l_{0}^{(i)}<l_{1}^{(i)}$
and $l_{n_i}^{(i)}<l'{}^{(i)}_{n_i+1}<l_{n_i+1}^{(i)}$ or cyclic.perm.}
\item \mbox{$n_i=0$,
$l_{0}'{}^{(i)}<l_{0}^{(i)}<l_{1}'{}^{(i)}<l_{1}^{(i)}$
or cyclic.perm.}
\item \mbox{$n_i=0$,
$l_{1}'{}^{(i)}<l_{0}^{(i)}<l_{0}'{}^{(i)}<l_{1}^{(i)}$
or cyclic perm.}
\end{enumerate}

We shall declare that two simple loops $C$ and $C'$ do not cross if and
only if
they have no crossing zone.

\medskip

We have the following important proposition:

\bp
If $C$ and $C'$ are  non crossing simple loops then the corresponding
Wilson
loop  $W^{\alpha}(C)$ and  $W^{\beta}(C')$ are commuting elements.
\ep
\smallskip

{\sl Proof:}

Let us  denote by ${\cal T}_{xyz}$
and ${\cal T'}_{xyz}$ the elements:

\be
{\cal T}^{\alpha\beta}_{xyz}=\left\{\begin{array}{ll}
 (\Rab_{12}\Rab_{21})^{-1}
 & \mbox{if $(x,y)<(y,z)$}\\
1\otimes 1 & \mbox{if $(x,y)>(y,z)$}
\end{array}\right.
\ee
and
\be
{\cal T'}^{\alpha\beta}_{xyz}=\left\{\begin{array}{ll}
 \Rab_{12}\Rab_{21}
 & \mbox{if $(x,y)>(y,z)$}\\
1\otimes 1 & \mbox{if $(x,y)<(y,z)$}
\end{array}\right.
\ee

It is easy to show the following relations using the basic commutation
rules in $\Lambda.$
Those are obtained by a tedious enumeration of all possible configurations
of
links and ciliations:

\begin{eqnarray*}
&&\ua(w,y)_1\ub(x,y)_2({\cal T}^{\alpha\beta}_{xyz})_{12}\ub(y,z)_2=
\ub(x,y)_2 \ub(y,z)_2\ua(w,y)_1 \\
&&\ua(y,w)_1\ub(x,y)_2\ub(y,z)_2 =
\ub(x,y)_2({\cal T}^{\alpha\beta}_{xyz})_{12} \ub(y,z)_2 \ua(y,w)_1\\
&&\mbox{if $(y,z)<(x,y)<(w,y)$ or cyclic perm.}\\
&&\ua(w,y)_1\ub(x,y)_2({\cal T'}^{\alpha\beta}_{xyz})_{12}\ub(y,z)_2=
\ub(x,y)_2 \ub(y,z)_2\ua(w,y)_1 \\
&&\ua(y,w)_1\ub(x,y)_2\ub(y,z)_2 =
\ub(x,y)_2({\cal T'}^{\alpha\beta}_{xyz})_{12} \ub(y,z)_2 \ua(y,w)_1\\
&&\mbox{if $(x,y)<(y,z)<(w,y)$ or cyclic perm.}\\
&&\ua(z,y)_1\ub(x,y)_2({\cal
T}^{\alpha\beta}_{xyz})_{12}\ub(y,z)_2\ub(z,w)_2 = \ub(x,y)_2
\ub(y,z)_2({\cal T}^{\alpha\beta}_{yzw})_{12}
\ub(z,w)_2\ua(z,y)_1\\
&&\ua(z,y)_1
\ub(x,y)_2({\cal T'}^{\alpha\beta}_{xyz})_{12}\ub(y,z)_2\ub(z,w)_2 =
\ub(x,y)_2 \ub(y,z)_2({\cal T'}^{\alpha\beta}_{yzw})_{12}\ub(z,w)_2
\ua(z,y)_1\\
&&\ua(y,z)_1\ub(x,y)_2\ub(y,z)_2
({\cal T}^{\alpha\beta}_{yzw})_{12}\ub(z,w)_2=\ub(x,y)_2
({\cal T}^{\alpha\beta}_{xyz})_{12}\ub(y,z)_{2}\ub(z,w)_2
\ua(y,z)_1\\
&&\ua(y,z)_1\ub(x,y)_2\ub(y,z)_2
({\cal T'}^{\alpha\beta}_{yzw})_{12} \ub(z,w)_2=\ub(x,y)_2
({\cal T'}^{\alpha\beta}_{xyz})_{12}\ub(y,z)_2
\ub(z,w)_2\ua(y,z)_1
\end{eqnarray*}

\medskip

 If $C$ and $C'$ have no common vertices
 then $W^{\alpha}(C)$ and  $W^{\beta}(C')$
are trivially commuting.

Let $C\cap C'$ be the set of common vertices of $C$ and $C'.$  $C\cap C'$
is naturally a disconnected union of paths
 $P_1,.., P_k$ (resp  $P_1',.., P_k'$ ) with the orientation induced by
 $C$ (resp. $C').$ For each of these paths we define
 $L_1,.., L_k$ (resp  $L_1',.., L_k'$ ) the paths defined as before.

It is sufficient to show that $M^{\alpha}(L_j)$ commute with $M^{\beta}(L_j')$
to prove the commutation relation between Wilson loops attached to $C$ and
$C'$ unless the marginal
case  where one of the $P_j$ or $P_j'$ is a closed curve (because in this
case one
of the $L_j$ or $L_j'$ is ill defined). This  case is studied at the end of
the proof.

Let us now consider a connected component of $C\cap C'$ and its related
paths
$L$ and $L'.$ Let us first suppose that $L_j$ and $L_j'$ have opposite
orientation.
we can write $L=(x_{n+1}, x_{n} \cdots x_{1}, x_{0})$
and the respective $L'=(x_{0}', x_1 \cdots, x_{n}, x_{n+1}').$
The fact that $C$ and $C'$ do not cross implies that
we fullfill one of the following condition:
\begin{eqnarray*}
&&\mbox{$(x_{0}',x_1)<(x_{0},x_1)<(x_1,x_2)$ and
$(x_{n-1},x_n)<(x_{n},x_{n+1})<(x_{n},x_{n+1}')$}\\
&&\mbox{or}\\
&&\mbox{$(x_{0},x_1)<(x_{0}',x_1)<(x_1,x_2)$ and
$(x_{n-1},x_n)<(x_{n},x_{n+1}')<(x_{n},x_{n+1})$}\\
&&\mbox{or cyclic perm.}
\end{eqnarray*}
Using the latter commutation properties and the latter remark, it is easy
to show that
$M^{\alpha}(L)$ commute with $M^{\beta}(L').$
This is more or less the standard
``railway proof'' of integrable models. In one of the latter situations
we have for example:
\begin{eqnarray*}
&&M^{\alpha}(L)_1 M^{\beta}(L')_2=\\
&=&\ua(x_{n+1},x_n)_1.\cdots
\ua(x_{1},x_0)_1\ub(x_{0}',x_1)_2\cdots\ub(x_{n},x_{n+1}')_2 \\
&=&\ua(x_{n+1},x_n)_1\cdots\ua(x_{2},x_1)_1
\ub(x_{0}',x_1)_2({\cal T}^{\alpha\beta}_{x_{0}'x_{1}x_{2}})_{12}
\ub(x_{1},x_2)_2\\
&&\hskip 40pt\cdots\ub(x_{n},x_{n+1}')_2\ua(x_{1},x_0)_1\\
&=&\ua(x_{n+1},x_n)_1
\ub(x_{0}',x_1)_2..\ub(x_{n-1},x_n)_{2}
{({\cal T}^{\alpha\beta}_{x_{n-1}x_{n}x_{n+1}'})}_{12}
\ua(x_{n},x_{n+1}')_2\\
& &\hskip 40pt \ua(x_{n},x_{n-1})_1\cdots\ua(x_{1},x_0)_1\\
&=&\ub(x_{0}',x_1)_2..\ub(x_{n},x_{n+1}')_2\ua(x_{n+1},x_n)_1\cdots\ua(x_{1}
,x_0)_1 \\
&=& M^{\beta}(L')_2 M^{\alpha}(L)_1
\end{eqnarray*}
If $L$ and $L'$ have the same orientation, we can easily modify the proof
to the result:
\begin{eqnarray*}
&&M^{\alpha}(L)_1 M^{\beta}(L')_2=\\
&=&\ua(x_{0},x_1)_1.\cdots
\ua(x_{n},x_{n+1})_1\ub(x_{0}',x_1)_2\cdots\ub(x_{n},x_{n+1}')_2 =\\
&=&\ua(x_{0},x_1)_1\cdots\ua(x_{n-1},x_n)_1
\ub(x_{0}',x_1)_2
\cdots\\
&&\hskip 40pt\cdots\ub(x_{n-1},x_{n})_2
({\cal T}^{\alpha\beta}_{x_{n-1}x_{n}x_{n+1}'})_{12}\ub(x_{n},x_{n+1}')_2
\ua(x_{n},x_{n+1})_1=\\
&=&\ua(x_{n+1},x_n)_1
\ub(x_{0}',x_1)_2{({\cal T}^{\alpha\beta}_{x_{0}'x_{1}x_{2}})}_{12}
\ub(x_{1},x_2)_2..\ub(x_{n-1},x_n)_{2}
\ua(x_{1},x_{2})_1..\ua(x_{n},x_{n-1})_1\\
&=&\ub(x_{0}',x_1)_2..\ub(x_{n},x_{n+1}')_2\ua(x_{0},x_1)_1\cdots\ua(x_{n},x
_{n+1})_1 \\
&=& M^{\beta}(L')_2 M^{\alpha}(L)_1
\end{eqnarray*}
The latter proofs do not work if exceptionally $x_0=x_{n},x_{n+1}=x_1$. In
this case
the proof is a bit different:

\begin{eqnarray*}
&&M^{\alpha}(L)_1 M^{\beta}(L')_2=\\
&=&\ua(x_{n+1},x_n)_1.\cdots \ua(x_{1},x_0)_1
\ub(x_{1},x_2)_2\cdots\ub(x_{n},x_{1})_2 =\\
&=&\ua(x_{n+1},x_n)_1\cdots\ua(x_{2},x_1)_1
{\cal R}_{12}(x_{0}x_{1}x_{2})\ub(x_{1},x_2)_2\cdots
\ub(x_{n},x_1)_2 {\cal R}_{12}(x_0x_{1}x_{n})^{-1}\times\\
&&\hskip 40pt\ua(x_{1},x_0)_1=\\
&=&\ua(x_{n+1},x_n)_1\cdots\ua(x_{3},x_2)_1
\ub(x_{1},x_2)_2
({\cal T}^{\alpha\beta}_{x_{1}x_{2}x_{3}})_{12}\ub(x_{2},x_3)_2\\
&&\cdots \ub(x_{n},x_1)_{2}\ua(x_{2},x_{1})_1\ua(x_{1},x_{0})_1\\
&=&\ua(x_{n+1},x_n)_1
\ub(x_{1},x_2)_2\cdots
\ub(x_{n-1},x_{n})_2({\cal T}^{\alpha\beta}_{x_{n-1}x_{n}x_{1}})_{12}
\ub(x_{n},x_1)_{2}\times \\
&& \hskip 40pt\ua(x_{n},x_{n-1})_1\cdots\ua(x_{1},x_{0})_1\\
&=&\ub(x_{1},x_2)_2..\ub(x_{n},x_{1})_2\ua(x_{n+1},x_n)_1\cdots\ua(x_{1},x_0
)_1 \\
&=& M^{\beta}(L')_2 M^{\alpha}(L)_1.
\end{eqnarray*}

In order to complete the proof of this theorem it remains to study the case
where the two loops
$C$ and $C'$ are equal up to orientation. Let
$C=(x_1,\cdots,x_n,x_{n+1}=x_1)$ it is   easy to show using an
induction proof on $k\leq n+1$ that we have:
\bea
\Rab_{12}M^{\alpha}(x_2,\cdots,x_k)_1
M^{\beta}(x_2,\cdots,x_k)_2 &=&M^{\beta}(x_2,\cdots,x_k)_2
M^{\alpha}(x_2,\cdots,x_k)_1\Rab_{21}
\nonumber\\
\Rab_{21}^{-1}M^{\alpha}(x_2,\cdots,x_k)_1
M^{\beta}(x_2,\cdots,x_k)_2 &=&M^{\beta}(x_2,\cdots,x_k)_2
M^{\alpha}(x_2,\cdots,x_k)_1
\Rab_{12}^{-1}\nonumber.
\eea
As a result we get:
\bea
 & &{\cal R}^{\alpha\beta
-1}(x_1,x_2,x_3)_{12}M^{\alpha}(x_1,\cdots,x_{n+1})_1
{\cal R}^{\alpha\beta}(x_n,x_1,x_2)_{12}M^{\beta}(x_1,\cdots,x_k)_2
=\nonumber\\
&=& {\cal R}^{\alpha\beta
-1}(x_1,x_2,x_3)_{12}M^{\alpha}(x_1,\cdots,x_{n})_1\ub(x_1,x_2)_2
\ua(x_n,x_1)_1 M^{\beta}(x_2,\cdots,x_{n+1})_2\nonumber\\
&=& {\cal R}^{\alpha\beta -1}(x_1,x_2,x_3)_{12}\ua(x_1,x_2)_1
\ub(x_1,x_2)_2
{\cal R}^{\alpha\beta }(x_1,x_2,x_3)_{21}\times\nonumber\\
& &M^{\alpha}(x_2,\cdots,x_{n+1})_1
 M^{\beta}(x_2,\cdots,x_{n+1})_2\nonumber\\
&=& \ub(x_1,x_2)_2 \ua(x_1,x_2)_1
M^{\alpha}(x_2,\cdots,x_{n+1})_1
 M^{\beta}(x_2,\cdots,x_{n+1})_2\nonumber\\
&=&\ub(x_1,x_2)_2 \ua(x_1,x_2)_1 {\cal R}^{\alpha\beta }(x_1,x_2,x_3)_{12}
M^{\beta}(x_2,\cdots,x_{n+1})_2\times\nonumber\\
& &M^{\beta}(x_2,\cdots,x_{n+1})_1
{\cal R}^{\alpha\beta -1}(x_1,x_2,x_3)_{21}\nonumber\\
&=& M^{\beta}(x_1,\cdots,x_{n})_2\ua(x_1,x_2)_1\ub(x_n,x_1)_2
M^{\alpha}(x_2,\cdots,x_{n+1})_1
{\cal R}^{\alpha\beta -1}(x_1,x_2,x_3)_{21}\nonumber\\
&=&M^{\beta}(x_1,\cdots,x_{n+1})_2
{\cal R}^{\alpha\beta}(x_n,x_1,x_2)_{12}M^{\alpha}(x_1,\cdots,x_k)_1
{\cal R}^{\alpha\beta -1}(x_1,x_2,x_3)_{21}\nonumber.
\eea

Applying the permutation operator to this identity we obtain the equivalent
equation:
\bea
& &{\cal R}^{\alpha\beta
}(x_1,x_2,x_3)_{21}M^{\alpha}(x_1,\cdots,x_{n+1})_1
{\cal R}^{\alpha\beta}(x_n,x_1,x_2)_{12}M^{\beta}(x_1,\cdots,x_k)_2
=\nonumber\\
&=&M^{\beta}(x_1,\cdots,x_{n+1})_2
{\cal R}^{\alpha\beta}(x_n,x_1,x_2)_{12}M^{\alpha}(x_1,\cdots,x_k)_1
{\cal R}^{\alpha\beta }(x_1,x_2,x_3)_{12}\nonumber.
\eea

These two equivalent relations can be recast in
 the reflection equation:
\bea
& &{\cal R}_{12}^{\alpha\beta -1}(x_n,x_1,x_2)M^{\alpha}(C)_1
{\cal R}_{12}^{\alpha\beta }(x_n,x_1,x_2)M^{\beta}(C)_2=\nonumber\\
& &M^{\beta}(C)_2{\cal R}_{21}^{\alpha\beta }(x_n,x_1,x_2)M^{\alpha}(C)_1
{\cal R}_{12}^{\alpha\beta -1}(x_n,x_1,x_2).
\eea
Let us denote ${\cal R}^{-1}(x_n,x_1,x_2)=\sum_{i}a_i\otimes b_i,$ from
quasitriangularity properties we have:
\be
\sum_{i,j}a_i a_j\otimes S^{-2}(b_j) S^{-1}(b_i)=1\otimes 1.
\ee
Therefore:
\begin{eqnarray*}
& &tr_{V_{\alpha}}(\mua M^{\alpha}(x_1,\cdots,x_{n+1}))
tr_{V_{\beta}}(\mub M^{\beta}(x_1,\cdots,x_{n+1}))=\\
&=&\sum_{i,j}tr_{12}(\mua_1 \mub_2 M^{\alpha}(x_1,\cdots,x_{n+1})_1
a_i a_j\otimes S^{-2}(b_j) S^{-1}(b_i)M^{\beta}(x_1,\cdots,x_{n+1})_2)\\
&=&tr_{12}(\mua_1 \mub_2 {\cal R}_{12}^{\alpha\beta
-1}(x_n,x_1,x_2)M^{\alpha}(C)_1
{\cal R}_{12}^{\alpha\beta }(x_n,x_1,x_2)M^{\beta}(C)_2)\\
&=&tr_{12}(\mua_1 \mub_2 M^{\beta}(C)_2{\cal R}_{21}^{\alpha\beta
}(x_n,x_1,x_2)M^{\alpha}(C)_1
{\cal R}_{12}^{\alpha\beta -1}(x_n,x_1,x_2)\\
&=&tr_{V_{\beta}}(\mub M^{\beta}(x_1,\cdots,x_{n+1}))tr_{V_{\alpha}}(\mua
M^{\alpha}(x_1,\cdots,x_{n+1})).
\end{eqnarray*}
When $C'={bar C}$ a similar proof applies as well.

This ends the proof of the theorem.

$\Box$
\smallskip

All what we have previously done is a study of  a q-analog of
the space of configurations of
lattice gauge theory and its gauge covariance. We have not included for the
present time
the  quantum   fluctuations of
the gauge fields. This will be achieved in the next section.
\medskip

\section{Quantum q-deformed lattice gauge theory}

In ordinary lattice gauge field theory, integration over the fields uses
as a central tool the Haar measure. This has been already recalled in
Section 1. We will now
study a q-analog of this notion, i.e we will study  integrals over
$\Lambda$ invariant under
the coaction of the gauge algebra $\hat \Gamma.$

\bp[Invariant measure]
There exists a unique linear form $h:\Lambda\rightarrow \Lambda_{\partial
\Sigma}$ such that:
\begin{enumerate}
\item (invariance) $(h\otimes id)\Omega(A)= h(A)\otimes 1 \,\,\forall A\in
\Lambda$
\item (factorisation) $h(AB)=h(A)h(B)\\
\forall A\in\Lambda_X, \forall B\in\Lambda_Y,\forall X, Y\subset L,\,\,
(X\cup {\bar X}) \cap (Y\cup {\bar Y})=\emptyset$
\item $h_{\vert \partial \Sigma}=id_{\partial \Sigma}$
  \end{enumerate}
It can be evaluated on any element using the formula:
\be
h(\ua(x,y){}^i_j)=\delta_{\alpha,0}
\ee
where $0$ denotes the trivial representation of dimension $1,$ i.e $0$ is
the counit, and $(x,y)$ is an
interior edge.

It can be recursively computed on any element of $\Lambda$ using the
formula
\be
h(A\ua(x,y){}^i_jB)=\delta_{\alpha,0}h(AB) \label{intpartiel}
\ee
with $(x,y)$ an interior edge and   $A,B\in
\Lambda_X,\,(x,y)\notin\Lambda_X, \,(y,x)\notin\Lambda_X$

\ep

{\sl Proof}:

 From the assumed  invariance of $h$ we get that:
\be
h(\ua(x,y){}^i_j) 1\otimes1=(\ga{}^{x})^{i}_m
h(\ua(x,y){}^m_n)S((\ga{}^{y})^{n}_j )
\ee
Using the independance of the vectors $(\ga{}^{x})^{i}_m$  we get that
$h(\ua(x,y){}^i_j)=0,$ except in the case $\alpha=0.$
Let $M=L^b\cup\{l_1,\cdots, l_p\}$ be a subset of links such that $M$
contains the boundary links and each of the interior links in one and only
one orientation.
 From the assumption on the basis of $\Lambda$ we can write any element $A$
of
$\Lambda$ as a unique linear combination:
\be
A=\sum_{\alpha_1,\cdots, \alpha_p}
U_{L^b}^{\alpha_1,\cdots, \alpha_p}tr(a_{\alpha_1,\cdots,\alpha_p}
\prod_{j=1}^p \Ua{j}(l_j)_j)
\ee
where
$a_{\alpha_1,\cdots,\alpha_p}\in End(\bigotimes_{j=1}^p V_{\alpha_j})$ and
$U_{L^b}^{\alpha_1,\cdots, \alpha_p}\in \Lambda_{\partial \Sigma}.$

 From $h_{\vert \partial \Sigma}=id_{\partial \Sigma}$ and the assumed
factorisation property we get that $h(A)=a_{0,\cdots ,0}U_{L^b}^{0, \cdots,
0}.$
This shows uniqueness of   $h.$

It is straightforward  to show that $h$ defined by the last formula is
 invariant under the coaction of $\hat\Gamma$.

The factorisation property is proved using the  commutation
relations in $\Lambda$ and the relation
${\buildrel 0\alpha \over R}={\buildrel \alpha 0\over R}=id_{V_{\alpha}}.$
Relation (\ref{intpartiel}) is proved similarly.

$\Box$

\smallskip

It will be convenient in the rest of this article to use the notation $\int
dh$ instead of $h.$

\bp
Let (x,y) be an interior link, we have the important formula:
\be
h(\ua(y,x){}^i_j\ua(x,y){}^k_l)={1\over [d_{\alpha}]}(\mua^{-1})^k_j
\delta^i_l.
\ee
which can also be written:
\be
h(\ua(y,x)_1\mua_2 \ua(x,y)_2)={1\over [d_{\alpha}]}P_{12}.
\label{ortho}
\ee
\ep

{\sl Proof:}

Let $(x,y)$ be an interior edge  from the decomposition
rule we get:
\be
h(\uabar(x,y)_1\ua(x,y)_2)=
\phi_{\bar{\alpha},\alpha}^{0}\psi^{\alpha,\bar{\alpha}}_{0}P_{12}=
\psi^{\bar{\alpha},\alpha}_{0}\Rabara_{21}v_{\alpha}^{-1}.
\ee
Using the known expressions of the right handside (see proposition (1)) we
 obtain:
\be
h(\uabar(x,y)^{i}_j\ua(x,y)^k_l)={1\over [d_{\alpha}]}
(\sua{}^{-1})^k_i \sua{}^j_l.
\ee
Using relation (\ref{ul=ubarl'1}) we finally prove the relation
(\ref{ortho}).

$\Box$

\smallskip

We will now define an analog of the Yang-Mills action which is, as usual,
built with elementary Wilson
loops attached to faces of the triangulation.

\bd[Boltzmann weights]
Let $F$ be a face of the triangulation, we  define a Boltzmann weight
element
of $\Lambda$:
\be
W_F=\sum_{\alpha\in Irr(A)}[d_{\alpha}]W^{\alpha}(\partial F)e^{-a_F
C_{\alpha}}
\ee
where $a_F$ is the area of the face $F$ and $C_{\alpha}$ are fixed non
negative
numbers.
\ed

This element is the non commutative analog of the Boltzmann weight
 (\ref{wabeta}).
If $F$ and $F'$ are two faces, proposition (3) implies that
$W_F$ and $W_{F'}$ commute. We can then define unambiguously the gauge
invariant element:
$\prod_{i=1}^{n_F}W_{F_i}$ which does not depend on the labelling of faces.

\bp[Yang Mills measure]
The Yang Mills measure $h_{YM}$ on $\Lambda$ is defined as follows:
\be
h_{YM}(A)=h(A\prod_{i=1}^{n_F}W_{F_i})\,\,\,\,\forall A\in \Lambda.
\ee
It is an invariant measure in the sense that:
\be
(h_{YM}\otimes id)\Omega(A)=h_{YM}(A)\otimes 1\,\,\forall A\in \Lambda
\ee
\ep

It will be convenient to use the notation:

\be
h_{YM}(a)=\int_{\Sigma} a\,\, dh_{YM}.
\ee

{\sl Proof}:
Let $A$ be an element of $\Lambda,$
We have:
\begin{eqnarray*}
(h_{YM}\otimes id)\Omega(A)&=&(h\otimes id)(\Omega(A)
\prod_{i=1}^{n_F}W_{F_i}\otimes 1)\\
&=&(h\otimes id)(\Omega(A)\Omega(\prod_{i=1}^{n_F}W_{F_i}))\\
&=&(h\otimes id)(\Omega(A\prod_{i=1}^{n_F}W_{F_i})=h_{YM}(A)\otimes 1.
\end{eqnarray*}
$\Box$

\bd[Correlation  functions]
Let $\Sigma$ be a triangulated Riemann surface with boundary
$\partial \Sigma=\bigcup_{i=1}^{n}C_i$ where $C_i$ are nonintersecting
simple loops.
We define the  partition function $Z(\Sigma;C_1,\cdots,C_n)$ to be the
element of
$\bigotimes_{i=1}^{n}\Lambda_{C_i}$ defined by:
\be
Z(\Sigma;C_1,\cdots,C_n) =\int_{\Sigma} 1\,\, dh_{YM}.
\ee
\ed

\bp[Locality]
Let  $\Sigma$ be the same surface as before, and let us consider $C$ a
simple loop (consisting
in links belonging to the triangulation) which divides
$\Sigma$ in two pieces $\Sigma_1$ and $\Sigma_2.$

We have the usual Markov (or locality ) property:
\be
Z(\Sigma;C_1,\cdots,C_n)=\int Z(\Sigma_1;C_1,\cdots,C_i,C)
Z(\Sigma_2;C_{i+1},\cdots,C_n,{\bar C})
\prod_{l\in C} dh(l).
\ee

\ep

{\sl Proof:}

 Commutation of Boltzmann weigths implies obviously Markov property.
$\Box$

\medskip

We will later on use  this
 locality property to compute any correlation
functions using only two and three points correlation functions.

\medskip

In the rest of this work we will choose a precise expression (see
proposition 8)
for the functions
$\omega_{\alpha}(C).$ This will implies that the theory is
quasitopological.
In order to simplify expressions in the proofs we will moreover assume
that $W_{F}$ is independant of the area
of the face, i.e we are choosing $C_{\alpha}=0.$  We could as well
compute the correlation
functions of the continuum limit of the theory in the case where
$C_{\alpha}\not=0.$ The relationships between correlation functions of
the theory where $C_{\alpha}\not=0$ and the topological one is the same as
those
described by \cite{Ru}\cite{Wi} in the undeformed case.

\smallskip

Recall the theorem of Alexander \cite{G} in the two dimensional case:

let $K$, $L$ be compact simplicial complexes of dimension two and
$\Sigma_K,$
(resp. $\Sigma_L$) the topological surfaces defined by $K$ (resp. $L$),
$\Sigma_K$ is homeomorphic to $\Sigma_L$ if and only if $K$ and $L$ admit a
common subdivision obtained by a finite number of stellar moves.

The stellar move can be conveniently replaced by an equivalent set of two
moves called Matveev moves $M_1,M_2$ which consist respectively in:
\begin{enumerate}
\item  replacing any triplet of triangular faces described by their
boundary links:
$(x,a,y), ({\bar y},b,{\bar z}),
(z,c,{\bar x}) $ by
the triangular face $(a,b,c).$
\item  replacing a couple of neighbour triangular faces, described by their
boundary links,
$(a,b,x), ({\bar x},c,d)$
by
the couple $(a,y,d), ({\bar y},b,c).$
\end{enumerate}

\smallskip

It is important to remark that each move applied to the triangulation
leaves
the set of boundary links fixed.

In order to simplify the notations inside the following proofs, we have
made a
convenient abuse of notation and have used the same letter $l$ to denote
the link $l$
and the gauge variable $U_l$ attached to it.
We also have used the convention of labelling the
loops by their links and not by their vertices.

\smallskip

\bp[Topological invariance]
If $\omega_{\alpha}$ satisfies the following property:
\be
\omega_{\alpha}(C_1\#C_2)=\omega_{\alpha}(C_1)\omega_{\alpha}(C_2)
\ee
for any simple loops $C_1,C_2$ of length greater than three  with  common
edges,
then the
correlation functions do not depend on the triangulation of the surface
with
fixed boundaries.

A particular set of $\omega_{\alpha}$ is defined as follows:
let $t_{\alpha}$ be any complex numbers,
\be
\omega_{\alpha}(C)=
v_{\alpha}^{t_{\alpha}(1+{1\over 2}\sum_{x\in C}\epsilon(x,C))}.
\ee

\ep

{\sl Proof:}

Let us first prove  invariance under the first Matveev move:

Let us glue  three Boltzmann weights
associated to triangular faces labelled by their boundary
$C_1=(x,a,y), C_2=({\bar y},b,{\bar z}), C_3=(z,c,{\bar x})$
around a same vertex $m.$ The result after integration on the common links
is simply
the Boltzmann weight of the triangular plaquette
$C=(a,b,c).$ Indeed we have:
\begin{eqnarray*}
&&\int dh(x)dh(y)dh(z) W(x,a,y)W(y^{-1},b,z^{-1})W(z,c,x^{-1})=\\
&=&\int dh(x)dh(y)dh(z)\sum_{\alpha,\beta,\gamma}
[d_{\alpha}][d_{\beta}][d_{\gamma}]
 \rho_{\alpha}(m,C_1)\rho_{\beta}(m,C_2)\rho_{\gamma}(m,C_3)\times\\
&&\omega_{\alpha}(C_1)\omega_{\beta}(C_2)\omega_{\gamma}(C_3){\rm
tr}_{1}(\mua_1 \xa_1 \aa_1 \ya_1)
{\rm tr}_{2}(\mub_2 \yb\!{}^{-1}_2 \bb_2 \zb\!{}^{-1}_2)
{\rm tr}_{3}(\muc_3 \zc_3 \cg_3 \xc\!{}^{-1}_3)
\end{eqnarray*}

but we have, using the integration formula (\ref{ortho}), that:
\begin{eqnarray*}
&&\int dh(x)dh(y)dh(z){\rm tr}_{123}(\mua_1 \xa_1 \aa_1 \ya_1
\mub_2 \yb\!{}^{-1}_2 \bb_2 \zb\!{}^{-1}_2
\muc_3 \zc_3 \cg_3 \xc\!{}^{-1}_3)\\
&=&\int dh(x){\rm tr}_{123}(\mua_1 \xa_1 \aa_1
\delta_{\alpha,\beta} [d_{\alpha}]^{-1}P_{12}
\bb_2 \delta_{\beta,\gamma}[d_{\beta}]^{-1}P_{23}
\cg_3 \xc\!{}^{-1}_3)=\\
&=&\delta_{\alpha,\beta,\gamma} {[d_{\alpha}]}^{-2}
\int dh(x){\rm tr}_{123}(P_{12}P_{23}\mua_3 \xa_3 \aa_3
\ba_3 \ca_3 \xa\!{}^{-1}_3)\\
&=&\delta_{\alpha,\beta,\gamma}{[d_{\alpha}]}^{-2}
\int dh(x){\rm tr}(\mua \xa\aa
\ba\ca \xa\!{}^{-1}).
\end{eqnarray*}

In order to integrate  over the link $x$, we can use  proposition (2)
and the relation
\be
\prod_{i=1}^3\rho_{\alpha}(m,C_i)=\rho_{\alpha}(m,(x,a,b,c,x^{-1})),
\ee
so that we can write:

\begin{eqnarray*}
&&\int dh(x)dh(y)dh(z) W(x,a,y)W(y^{-1},b,z^{-1})W(z,c,x^{-1})=\\
&=&\int dh(x)\sum_{\alpha} [d_{\alpha}]\omega_{\alpha}(C)
{\rm tr}_{1\cdot\cdot 5}(\mua{}^{\otimes 5} P_{54}\cdot\cdot P_{21}
\xa_1{\cal R}_{12}\aa_2{\cal R}_{23}
\ba_3{\cal R}_{34}\ca_4{\cal R}_{45}\xa{}^{-1}_5)\\
&=&\int dh(x)\sum_{\alpha} [d_{\alpha}]\omega_{\alpha}(C)
{\rm tr}_{1\cdot\cdot 5}(\mua_1 \mua_2 \mua_3 \mua_5
P_{54}\cdots P_{21}
\aa_2{\cal R}_{23}
\ba_3{\cal R}_{34}\xa_1 \mua_5 \xa\!{}^{-1}_5
\ca_4)\\
&=&\sum_{\alpha}\omega_{\alpha}(C)
{\rm tr}_{12345}(\mua_1 \mua_2 \mua_3 \mua_5
P_{15}P_{54}\cdots P_{21}
\aa_2{\cal R}_{23}
\ba_3{\cal R}_{34}
\ca_4)\\
&=&\sum_{\alpha} [d_{\alpha}]\omega_{\alpha}(C)
{\rm tr}_{234}( \mua_2 \mua_3 \mua_4
P_{43}P_{32}
\aa_2{\cal R}_{23}
\ba_3{\cal R}_{34}
\ca_4)\\
&=&W(a,b,c)
\end{eqnarray*}
which shows that $M_1$ is satisfied.

\smallskip

Let us now show the invariance under the second Matveev move.
Consider two triangles $C_1=(a,b,x), C_2=({\bar x},c,d)$ and denote by $m$
the vertex common to $a, x$ and $d.$

\begin{eqnarray*}
&&\int dh(x) W(a,b,x)W(x^{-1},c,d)=\\
&=&\int dh(x)\sum_{\alpha,\beta} [d_{\alpha}][d_{\beta}]
\rho{\alpha}(m,C_1)
\rho{\beta}(m,C_2)\omega_{\alpha}(C_1)\omega_{\beta}(C_2)
{\rm tr}_{12}(\mua_1 \aa_1 \ba_1
\xa_1 \mub_2 \xb\!{}^{-1}_2 \cb_2
\db_2) \\
&=&\sum_{\alpha,\beta}
[d_{\alpha}][d_{\beta}]\rho_{\alpha}(m,C)\omega_{\alpha}(C)
{\rm tr}_{12}(\mua_1 \aa_1 \ba_1
\delta_{\alpha,\beta}[d_{\alpha}]^{-1}P_{12} \cb_2
\db_2) \\
&=&\sum_{\alpha} [d_{\alpha}]
\rho_{\alpha}(m,C)\omega_{\alpha}(C){\rm tr}(\mua\aa \ba
\ca \da) \\
&=&W(a,b,c,d).
\end{eqnarray*}
We just have to use cyclicity of $W(a,b,c,d)$ and to inverse the previous
computation to conclude
the proof.

$\Box$

Remark. Let $C_1$ and $C_2$ be simple loops of length greater than three
with common edges, let $C_1\cap C_2$
be the set of common edges, then it is easy to show  the convolution
property:
\be
W(C_1\#C_2)=\int W(C_1)W(C_2)\prod_{l\in C_1\cap C_2} dh(l).
\ee

In the rest of this work we will assume that

\be
\omega_{\alpha}(C)=
v_{\alpha}^{t_{\alpha}(1+{1\over 2}\sum_{x\in C}\epsilon(x,C))}.
\ee

\bp[characters]
Let $C$ be a simple closed curve on $\Sigma$ we will define a ``character''

\be
\chia{}(C)=v_{\alpha}^{-1-t_{\alpha}}W_{\alpha}(C).
\ee
These elements satisfy the following usual orthonormality property:

\be
\int \chia{}(C)\chib{}({\bar C})\prod_{l\in C} dh(l) = \delta_{\alpha,
\beta}.
\ee

\ep

{\sl Proof:}

The orthogonality relation is proved as follows:
\begin{eqnarray*}
&&\int \chia{}(C)\chib{}({\bar C})\prod_{l\in C} dh(l)=\\
&=&\int v_{\alpha}^{-1-t_{\alpha}}v_{\beta}^{-1-t_{\beta}}
\rho_{\alpha}(C)\rho_{\beta}(C^{-1})
{\rm tr}_{12}(\mua_{1} \prod_{i=1}^k\ua(l_i)_{1}
\times\\
&&\hskip 40pt \omega_{\alpha}(C) \omega_{\beta}(C^{-1})
\mub_{2} \prod_{i=k}^1\ub(l_i')_{2})\prod_{l\in C} dh(l)\\
&=&\int {\rm tr}_{12}(\mua_{1} \prod_{i=1}^{k-1}\ua(l_i)_{1}
[d_{\alpha}]^{-1}\delta_{\alpha,\beta} P_{12} \prod_{i=k-1}^1\ub(l_i')_{2})
\prod_{i=1}^{k-1} dh(l_{i})\\
&=&\int \delta_{\alpha,\beta}[d_{\alpha}]^{-1}
{\rm tr}_{12}( P_{12}\mua_{1}  \prod_{i=1}^{k-1}\ua(l_i)_{1}
\prod_{i=k-1}^1\ua(l_i){}^{-1}_{1})
\prod_{i=1}^{k-1} dh(l_{i})\\
&=&\delta_{\alpha,\beta}
[d_{\alpha}]^{-1}tr_{12}(P_{12}\mua_{1})\\
&=&\delta_{\alpha,\beta}
\end{eqnarray*}

\medskip

\bp[Two and three points correlation functions]
Let $S_{2,2}$ be the two holed sphere with   boundary loops $C_1,C_2$ and
$S_{2,3}$ be the three holed sphere with   boundary loops $C_1,C_2,C_3$.
The expression of the two points and three points correlation function is
expressed
in term of characters as:
\bea
Z(S_{2,2};C_1,C_2)=\sum_{\alpha}\chia{}_{S_{2,2},C_1}\chia{}_{S_{2,2},C_2}
\label{twopoints}\\
Z(S_{2,3};C_1,C_2,C_3)=\sum_{\alpha}v_{\alpha}^{-1-t_{\alpha}}[d_{\alpha}]^{
-1}
\chia{}_{S_{2,3},C_1}\chia{}_{S_{2,3},C_2}\chia{}_{S_{2,3},C_3}\label{threep
oints}
\eea
\ep

{\sl Proof:}

Let us compute the two points correlation function with the loops $C_1$ and
$C_2$ as boundaries.

To simplify the proof we consider the cylinder obtained by
gluing two opposite edges of an hexagon $C=(a,b,c,d,e,{\bar c}).$ Let us
denote by $m$ (resp. $n$)
the departure point of $b$ (resp. $e$). We  have $C_1=(a,b)$ and
$C_2=(d,e).$

\begin{eqnarray*}
&&\int dh(c) W(a,b,c,d,e,c^{-1})=\\
&=&\int dh(c)\sum_{\alpha} [d_{\alpha}]
\rho_{\alpha}(m,C)\omega_{\alpha}(C){\rm tr}(\mua \ba \ca \da \ea
\ca\!{}^{-1} \aa) \\
&=&\int dh(c)\sum_{\alpha} [d_{\alpha}]v_{\alpha}\omega_{\alpha}(C)
{\rm tr}_{1\cdot\cdot 6}(\mua{}^{\otimes 6}
\sigma^{(6)}
\ba_1{\cal R}_{12}
\ca_2{\cal R}_{23} \da_3 {\cal R}_{34}
\ea_4{\cal R}_{45}\ca{}^{-1}_5{\cal R}_{56}\aa_6)\\
&=&\int dh(c)\sum_{\alpha} [d_{\alpha}]v_{\alpha}\omega_{\alpha}(C)
{\rm tr}_{1\cdot\cdot 6}(\mua_1 \mua_2\mua_3\mua_5\mua_6
\sigma^{(6)} \ba_1{\cal R}_{12}
\da_3 {\cal R}_{34}\ca_2 \mua_5\ca\!{}^{-1}_5
\ea_4{\cal R}_{56}\aa_6)\\
&=&\sum_{\alpha} v_{\alpha}\omega_{\alpha}(C)
{\rm tr}_{1\cdot\cdot 6}(\mua_1 \mua_2\mua_3\mua_5\mua_6
\sigma^{(6)} P_{25} \ba_1{\cal R}_{15}{\cal R}_{56}\aa_6
\da_3{\cal R}_{34}\ea_4).
\end{eqnarray*}

Using the relation $\sigma^{(6)}P_{25}=P_{43}P_{32}P_{65}P_{51},$
the previous equation can also be written:
\begin{eqnarray*}
&=&\sum_{\alpha} v_{\alpha}\omega_{\alpha}(C)
{\rm tr}_{156}(\mua_1 \mua_5 \mua_6
P_{65}P_{51} \ba_1{\cal R}_{15}{\cal R}_{56}\aa_6)
{\rm tr}_{34}(\mua_3 \mua_4 P_{34}\da_3
{\cal R}_{34}\ea_4)\\
&=&\sum_{\alpha}v_{\alpha}^{-2-2t_{\alpha}}\omega_{\alpha}(C_1)\omega_{\alpha}(C_2)
\rho_{\alpha}(m,C_1)\rho_{\alpha}(n,C_2){\rm tr}(\mua \ba \aa)
{\rm tr}(\mua \da \ea).
\end{eqnarray*}
Therefore we have proven (\ref{twopoints}) when $C_1, C_2$ have length two.

The structure of the proof is the same when $C_1,  C_2$ have different
lengths.

\medskip

Let us now compute the three points correlation function with the loops
$C_1,C_2,C_3$
as boundaries. To simplify the proof we consider the three holed sphere
obtained from
a decagon $C=(b^{(2)},a,c^{(1)},e,d^{(1)},d^{(2)},{\bar e},c^{(2)},{\bar
a},b^{(1)}).$
We will denote by $x,y,z$ respectively the departure points of
$b^{(1)},a^{(1)},c^{(1)}$.
\begin{eqnarray*}
&&\int dh(a)dh(e)
W(b^{(2)},a,c^{(1)},e,d^{(1)},d^{(2)},e^{-1},c^{(2)},a^{-1},b^{(1)})=\\
&=&\int dh(a)dh(e)\sum_{\alpha} [d_{\alpha}]
v_{\alpha}\omega_{\alpha}(C){\rm tr}_{0\cdot\cdot9}(\mua^{\otimes 10}
P_{09} \cdot\cdot P_{21}
\ba{}^{(2)}_1{\cal R}_{12}\aa_2{\cal R}_{23}\ca{}^{(1)}_3\times\\
&&{\cal R}_{34}\ea_4{\cal R}_{45} \da{}^{(1)}_5{\cal
R}_{56}\da{}^{(2)}_6{\cal R}_{67}\ea{}^{-1}_7
{\cal R}_{78} \ca{}^{(2)}_8{\cal R}_{89}
\aa_9{}^{-1}{\cal R}_{90} \ba{}^{(1)}_0) \\
&=&\int dh(a)dh(e)\sum_{\alpha} [d_{\alpha}]
v_{\alpha}\omega_{\alpha}(C){\rm tr}_{0\cdot\cdot9}
(\mua_{0}\mua_{9}\mua_{7}\mua_{5}\cdot\cdot\mua_{1}
P_{09}\cdot\cdot P_{21} \ba{}^{(2)}_1{\cal
R}_{12}\ca{}^{(1)}_3\times\\
&&{\cal R}_{34}
\da{}^{(1)}_5{\cal R}_{56}\ea_4\mua_{7}\ea{}^{-1}_7\da{}^{(2)}_6 {\cal
R}_{78}
\aa_2 \mua_{9}\aa_9{}^{-1}
\ca{}^{(2)}_8
{\cal R}_{90} \ba{}^{(1)}_0) \\
&=&\sum_{\alpha} [d_{\alpha}]^{-1}\omega_{\alpha}(C)
v_{\alpha}{\rm tr}_{0\cdot\cdot9}
(\mua_{0}\mua_{9}\mua_{7}\mua_{5}\cdot\cdot\mua_{1}
P_{09}\cdot\cdot P_{21}P_{47}P_{29} \ba{}^{(2)}_1{\cal
R}_{19}\ca{}^{(1)}_3\times\\
&&{\cal R}_{37}
\da{}^{(1)}_5{\cal R}_{56}\da{}^{(2)}_6 {\cal R}_{78}
\ca{}^{(2)}_8
{\cal R}_{90} \ba{}^{(1)}_0) \\
&=&\sum_{\alpha} [d_{\alpha}]^{-1}
v_{\alpha}\omega_{\alpha}(C){\rm tr}_{019}(\mua_{0}\mua_{1}\mua_{9}
P_{09}P_{91} \ba{}^{(2)}_1{\cal
R}_{19}{\cal R}_{90} \ba{}^{(1)}_0)\times\\
&&{\rm tr}_{2378}(\mua_{2}\mua_{3}\mua_{7}P_{87}P_{73}P_{32}
\ca{}^{(1)}_3{\cal R}_{37}{\cal R}_{78}\ca{}^{(2)}_8)
{\rm tr}_{456}(\mua_{4}\mua_{5}P_{65}P_{54}
\da{}^{(1)}_5{\cal R}_{56}\da{}^{(2)}_6 )\\
&=&\sum_{\alpha} [d_{\alpha}]^{-1}
v_{\alpha}^{-4-4t_{\alpha}}\prod_{i=1}^3\omega_{\alpha}(C_i)\rho_{\alpha}(C_
i)
{\rm tr}(\mua \ba{}^{(1)} \ba{}^{(2)}){\rm tr}(\mua \ca{}^{(1)}
\ca{}^{(2)})
{\rm tr}(\mua \da{}^{(1)}\da{}^{(2)}).
\end{eqnarray*}
Therefore we have proven (\ref{threepoints}) when $C_1, C_2, C_3$ have
length two.
The structure of the proof is the same when $C_1,  C_2, C_3$ have different
lengths.

$\Box$

\bp[Partition function of a general Riemann surface]

The partition function $Z_g$ of a compact connected orientable Riemann
surface of genus $g$
 without boundary is given by the formula:
\be
Z_g=\sum_{\alpha} (v_{\alpha}^{1+t_{\alpha}}[d_{\alpha}])^{2-2g}\label{Z}
\ee
\ep
{\sl Proof:}

this result is easily obtained as in the undeformed case by using the
expression
of the two and three points correlation functions and the orthogonality of
the characters \cite{Ru}\cite{Wi}.

$\Box$

Remark 1. When $C_{\alpha}\not=0$ the last formula is modified as follows:
\be
Z_g=\sum_{\alpha}
(v_{\alpha}^{1+t_{\alpha}}[d_{\alpha}])^{2-2g}e^{-A_{\Sigma} C_{\alpha}}
\ee
where $A_{\Sigma}$ is the area of the Riemann surface $\Sigma.$

Remark 2. The previous proposition clearly shows that the partition
function is
independant of the choice of the ciliation we choose to define the algebra
of
gauge fields.

\bp[Correlation functions]
The correlation functions

$Z_{g,n}(C_1,\cdots, C_n)$ of a Riemann surface of genus $g$ with boundary
$\partial \Sigma= \cup_{i} C_i$ is given
by the formula:
\be
Z_g=\sum_{\alpha}
(v_{\alpha}^{1+t_{\alpha}}[d_{\alpha}])^{2-2g-n}\prod_{i=1}^n
\chia{}_{\Sigma,C_i}
\ee
\ep

{\sl Proof:}

The  proof is the same as in the classical case. It suffices to cut the
surface
in three holed spheres glued along their boundaries and to use proposition
10.

$\Box$

Up to this point the choices of the $t_{\alpha}$ where completely
arbitrary.
 From propositions (11-12) we clearly see that the choice $t_{\alpha}=-1$ is

particularly important. In that case final formulas exhibit a complete
symmetry
in the exchange of q and $q^{-1}.$ If we take ${\cal G}=sl_2$ and  formally
set
$q$ being a root of unity and truncate the spectrum according to the value
of $q$ the partition function
$Z_g$ is equal to the Turaev-Viro invariant of $\Sigma\times [0,1].$ This
tends
to support the  connection between our theory at  $t_{\alpha}=-1$ with
hamiltonian
Chern-Simons theory.

In the rest of this work we will set $t_{\alpha}=-1$ and prove the fusion
identities for characters. We will moreover show that $W_{F_{i}}$ is
a projector.But it would be nice to introduce a new notation:
\bd
 From now the holonomy ${\cal H}$  associated to a simple closed curve
$C=(x_1,x_2,...,x_{n+1}=x_1)$ will be a renormalised version of the
previous definition:
\be
{\cal H}^{\alpha}(x_1,x_2,...,x_{n+1}=x_1)=
v_{\alpha}^{{1\over2}((\sum_{x\not=x_1}\epsilon(x,C)) - \epsilon(x_1,C))}
\prod_{i=1}^{n}\Ua {}(x_i,x_{i+1}).
\ee
then we have:
\be
\chi_{\alpha}(C)=tr(\mua{\cal H}^{\alpha}(C))
\ee
and for each plaquettes $F$:
\be
W_F=\sum_{\alpha}[d_{\alpha}]\chi_{\alpha}(\partial F)
\ee
\ed

\medskip

\bp[Fusion of characters]
Let $C$ be a simple loop, we have the following fusion identity on
characters:
\be
\chi_{\alpha}(C)\chi_{\beta}(C)=
\sum_{\gamma}N_{\alpha\beta}^{\gamma} \chi_{\gamma}(C).
\ee
It then follows that for each face $F$ of the triangulation $W_{F}$ is a
projector, i.e: $W_{F}^2=W_{F}.$
\ep

{\sl proof}
We will prove it when $C$ is of length three. The generalization to
arbitrary
length is straightforward. Let $C=(x,y,z)$, where $x, y, z$ are the links
of
$C$ where $x=(m,n), y=(n, p), z=(p,m),$ using the trick of the last part of

the proof of proposition (3) we can write:
\begin{eqnarray*}
&& \xa_1 \ya_1 \za_1 \xb_2 \yb_2 \zb_2= \\
&&=\sum_{i,j}\xa_1 \ya_1 \za_1 a_1^{(i)}a_1^{(j)}S^{-2}(b_{2}^{(j)})
S^{-1}(b_2^{(i)}) \xb_2 \yb_2 \zb_2\\
&&=\sum_{i,j}S^{-2}(b_{2}^{(j)})\xa_1 \ya_1 \za_1 {\cal R}_{12}(p,m,n)
\xb_2 \yb_2 \zb_2 a_1^{(j)}\\
&&=\sum_{i,j}S^{-2}(b_{2}^{(j)})\xa_1 \ya_1
\xb_2 \za_1 \yb_2 \zb_2 a_1^{(j)}\\
&&=\sum_{i,j}S^{-2}(b_{2}^{(j)})\xa_1
\xb_2{\cal R}_{21}(m,n,p)\ya_1 \yb_2  {\cal R}_{21}(n,p,m)\za_1 \zb_2
a_1^{(j)}\\
&&=\sum_{i,j\atop \gamma, m}S^{-2}(b_{2}^{(j)})\phi_{\alpha\beta}^{\gamma,
m}\xc
\yc \zc \psi^{\beta,\alpha}_{\gamma,m} P_{12} a_1^{(j)}
\lambda_{\alpha\beta\gamma}^{-\epsilon(n)-\epsilon(p)}.
\end{eqnarray*}

As a result we get:
\begin{eqnarray*}
&&tr_{12}(\mua_1 \mub_2 \xa_1 \ya_1 \za_1\xb_2 \yb_2 \zb_2)=
\sum_{i,j\atop \gamma, m}tr_{12}(\mua_1 \mub_2
S^{-2}(b_{2}^{(j)})\phi_{\alpha\beta}^{\gamma, m}\xc
\yc \zc \psi^{\beta,\alpha}_{\gamma,m} P_{12} a_1^{(j)})
\lambda_{\alpha\beta\gamma}^{-\epsilon(n)-\epsilon(p)}\\
&&\sum_{\gamma, m}tr_{12}(\mua_1 \mub_2\phi_{\alpha\beta}^{\gamma, m}\xc
\yc \zc \psi^{\beta,\alpha}_{\gamma,m} P_{12}{\cal R}_{12}^{-1}(p,m,n) )
\lambda_{\alpha\beta\gamma}^{-\epsilon(n)-\epsilon(p)}\\
&&\sum_{ \gamma}N_{\alpha\beta}^{\gamma}tr(\muc\xc \yc \zc)
\lambda_{\alpha\beta\gamma}^{-\epsilon(n)-\epsilon(p)+\epsilon(m)},
\end{eqnarray*}
which implies straightforwardly the fusion relation:
\be
\chi_{\alpha}(C)\chi_{\beta}(C)=
\sum_{\gamma}N_{\alpha\beta}^{\gamma} \chi_{\gamma}(C).
\ee
 From the relation $W_F=\sum_{\alpha}[d_{\alpha}]\chi_{\alpha}(\partial F)$
and
the fusion relation it follows that $W_F^2=W_F.$\\

$\Box.$

\medskip

Finally,it is possible to show that $W_F$ is in a certain sense
 a delta function which constraints the holonomy of the plaquettes to be
 equal to one.

\bp[delta function and flat connections]
The Boltzmann weight is a delta function located at the unit element
for the holonomy around the plaquette $C$:
\be
W_C \times ({\cal H}^{\beta}(C) - 1) = 0.
\ee
\ep

{\sl proof:}

In fact we will show the latter property in a slicely different form:
\be
W_C (tr(\mub{\cal H}^{\beta}(C) V) - tr(\mub V)) = 0
\ee
for any $V$ in $End(V_{\alpha}).$

With the same method as before we have:
\begin{eqnarray*}
W_F tr(\mub{\cal H}^{\beta}(C) V^{\beta})&=&
(\sum_{\alpha}[d_{\alpha}]\chi_{\alpha}(C)) tr(\mub{\cal H}^{\beta}(C)
V^{\beta})\\
&=&\sum_{\alpha}[d_{\alpha}]tr_{12}(\mua_1{\cal H}^{\alpha}(C)_1
\mub_2{\cal H}^{\beta}(C)_2 V^{\beta}_2)\\
&=&\sum_{\gamma,m}tr_{V_{\gamma}}(\muc{\cal H}^{\gamma}(C)
(\sum_{\alpha}[d_{\alpha}] \psi^{\beta\alpha}_{\gamma}
V^{\beta}\phi_{\beta\alpha}^{\gamma})).
\end{eqnarray*}
In order to simplify the last part of the latter equation, we must
establish some relations between Clebsch-Gordan coefficients.
There exist complex numbers $ A_{\gamma}^{\beta\alpha}$ and
$ B^{\gamma}_{\beta\alpha}$ such that:
\begin{eqnarray*}
&&\psi^{\beta,\alpha}_{\gamma}=A_{\gamma}^{\beta\alpha}
(\psi^{\beta,\bar{\beta}}_{0}\otimes id_{V_{\gamma}})
(id_{V_{\beta}}\otimes\psi^{\bar{\beta},\alpha}_{\gamma})\\
&&\phi_{\beta,\alpha}^{\gamma}=B^{\gamma}_{\beta\alpha}
(id_{V_{\beta}}\otimes\psi_{\bar{\beta},\alpha}^{\gamma})
(\phi_{\beta,\bar{\beta}}^{0}\otimes id_{V_{\gamma}})
\end{eqnarray*}
with :
\be
 A_{\gamma}^{\beta\alpha}
B^{\alpha}_{\bar{\beta},\gamma}=[d_{\beta}]\nonumber.
\ee
Using the scalar product it is also easy to obtain the relations:
\be
 B^{\alpha}_{\beta,\gamma}=[d_{\alpha}]
B^{\bar{\gamma}}_{\bar{\alpha},\beta}\nonumber
\ee
\be
 A^{\gamma}_{\beta,\alpha}=[d_{\gamma}]
B^{\bar{\gamma}}_{\bar{\alpha},\bar{\beta}}\nonumber.
\ee
The latter formulas give the transform:
\begin{eqnarray*}
& &\psi^{\beta,\alpha}_{\gamma}
V^{\beta}V'^{\alpha}\phi_{\beta,\alpha}^{\gamma}= \\
&=&[d_{\alpha}]^{-1}[d_{\beta}][d_{\gamma}]
(\psi^{\beta,\bar{\beta}}_{0}\otimes id_{V_{\gamma}})
(id_{V_{\beta}}\otimes\psi^{\bar{\beta},\alpha}_{\gamma})
V^{\beta}V'^{\alpha}
(id_{V_{\beta}}\otimes\psi_{\bar{\beta},\alpha}^{\gamma})
(\phi_{\beta,\bar{\beta}}^{0}\otimes id_{V_{\gamma}}).
\end{eqnarray*}
We are now able to conclude the proof:
\begin{eqnarray*}
& &\sum_{\alpha}[d_{\alpha}] \psi^{\beta,\alpha}_{\gamma}
V^{\beta}\phi_{\beta,\alpha}^{\gamma}= \\
&=&\sum_{\alpha,\gamma,m}[d_{\beta}][d_{\gamma}]
(\psi^{\beta,\bar{\beta}}_{0}\otimes id_{V_{\gamma}})
(id_{V_{\beta}}\otimes\psi^{\bar{\beta},\alpha}_{\gamma})
V^{\beta}
(id_{V_{\beta}}\otimes\psi_{\bar{\beta},\alpha}^{\gamma})
(\phi_{\beta,\bar{\beta}}^{0}\otimes id_{V_{\gamma}})\\
&=&[d_{\gamma}] id_{V_{\gamma}}
([d_{\beta}]\psi^{\beta,\bar{\beta}}_{0}V^{\beta}\phi_{\beta,\bar{\beta}}^{0
})\\
&=&[d_{\gamma}] id_{V_{\gamma}}
tr_{V_{\beta}}(\mub V^{\beta}).
\end{eqnarray*}

$\Box.$

\medskip

This property is essential
in the context of quantum gauge theory as described before \cite{Mi},
but it also allows us to understand why this theory is related
to Chern-Simons theory.

In our next work \cite{BRR} we will study more deeply the algebra of gauge
invariants elements and develop the properties of $P=\prod_{i}W_{F_i}$.
We will define gauge invariant and cyclic invariant objects associated
to intersecting curves and will show that the expectation values of these
Wilson
loops are in fact 3-dimensional knot invariants related to
Reshetikhin-Turaev
ribbon invariants in the three manifold $\Sigma\times[0,1].$

\section{Conclusion}

In this work we have defined and studied a two dimensional quantum gauge
field
theory  where the gauge symmetry is described by a quantum group. This
model is a
quasitopological field theory with an infinite number of fields.
We still have to understand the situation where $q$ is a root of unity
(this was not allowed in our setting). In order to solve this problem one
should
certainly use the formalism of weak quasi Hopf algebras \cite{MS} to ensure

truncation of the spectrum. This study has already been started in
\cite{AGS}.
 Our formalism do not use any
involution but it is highly desirable to introduce one to ensure reality
properties of the theory. It seems that when $q$ is real there is no
problem of
this type and the corresponding theory is a deformation of two dimensional
Yang Mills theory associated to  compact classical Lie groups.

This gauge theory appears to describe a three dimensional topological
theory,
i.e. that all correlation functions of Chern-Simons theory
can be obtained by a finite dimensional path integral formalism
in a two dimensional space time \cite{BRR}.
One of the remaining challenging problem is to extend the present formalism

to higher dimensional space time where the corresponding theories should be

far more interesting. In two dimension the definition of $\Lambda$ uses as
a central tool a cilium order at each vertex. This is easy to define using
the natural cyclic order on a two dimensional oriented surface. In higher
dimension it seems that there is no problem of definition of $\Lambda$
using
as well a cilium at each vertex but one has to  find clever permutation
invariant
objects in order to compensate non commutation of Boltzmann weights.

Another important problem which still remains open is to find
representations of
the gauge fields algebra $\Lambda.$
Representations of lattice Kac Moody algebras have already been found in
\cite{BB}\cite{AFS}. Constructions inspired by these works should lead to
representations of $\Lambda.$ Representations of this algebra is an
important
step to understand the precise continuum limit of those theories and to
compare
them with Hamiltonian Chern-Simons theory.

\medskip

{\bf Acknowledgements}

We would like to thank M.Petropoulos for interesting discussions at the
beginning of this work as well as
A.Yu.Alekseev, B.Enriquez, L.Freidel, J.M.Maillet, V.Pasquier,
N.Reshetikhin and
V.Schomerus for discussions and encouragements.

\bibliographystyle{unsrt}

\begin{thebibliography}{10}

\bibitem{AGS}
A.Y.Alekseev, H.Grosse, V.Schomerus,
\newblock{Combinatorial Quantization of the Hamiltonian Chern-Simons
Theory,}
\newblock{\it hep-th /94/03,} {\bf } (1994).


\bibitem{AFS}
A.Y.Alekseev, L.D.Faddeev, M.A.Semenov-Tian-Shansky,
\newblock{Hidden Quantum groups inside Kac-Moody algebra,}
\newblock{\it Comm.Math.Phys,} {\bf 149,} 335 (1992).





\bibitem{BB}
O.Babelon, L.Bonora,
\newblock{Quantum Toda theory,}
\newblock{\it Phys.Lett.,} {\bf B253}, 365 (1991).




\bibitem{Bo}
D.V.Boulatov,
\newblock{q-Deformed lattice gauge theory and three manifold invariants,}
\newblock{\it Int.J.Mod Phys,} {\bf A.8 }, 3139 (1993).


\bibitem{BM}
T.Brzezi\'nski, S.Majid,
\newblock{Quantum group gauge theory on quantum spaces,}
\newblock{\it Comm.Math.Phys,} {\bf 157}, 591  (1993).

\bibitem{BRR}
E.Buffenoir, N.Yu.Reshetikhin, Ph.Roche
\newblock{(in preparation)}


\bibitem{Cr}
M.Creutz,
\newblock{Quarks, gluons and lattices,}
\newblock{\it Cambridge University Press,} {\bf } (1983).


\bibitem{D}
V.G.Drinfeld,
\newblock{On almost cocomutative Hopf algebras,}
\newblock{\it Leningrad.Math.Journal,} {\bf 1}, 321 (1990).



\bibitem{FR}
V.V.Fock, A.A.Rosly,
\newblock{Poisson structure on moduli of flat connections on Riemann
surfaces and
r-matrices,} \newblock{\it Preprint ITEP 72-92,} {\bf } (1992).


\bibitem{G}
L.C.Glaser,
\newblock{Geometrical Combinatorial Topology,}
\newblock{\it Van Nostrand Reinhold Mathematical Study,} {\bf 27}, (1970).


\bibitem{Ji}
M.Jimbo, Editor,
\newblock{Yang Baxter equation in integrable systems,}
\newblock{\it Advances in Mathematical Physics,} {\bf Vol 10}, (1989).

\bibitem{KS}
M.Karowski, R.Schrader,
\newblock{A Combinatorial Approach to Topological Quantum Field Theories
and Invariants of Graphs,}
\newblock{\it Comm.Math.Phys,} {\bf 151,}  355 (1993).



\bibitem{MS}
G.Mack, V.Schomerus,
\newblock{Quasi quantum group symmetry and local braid relations
in the conformal Ising Model,}
\newblock{\it Phys.Lett.B,} {\bf 267 }, 207 (1991).


\bibitem{Mi}
A.A.Migdal,
\newblock{Recursion equations in gauge field theories,}
\newblock{\it Sov.Phys.JETP,} {\bf 42}, 413 (1975).


\bibitem{FRT}
N.Yu.Reshetikhin, L.A.Takhtajan and L.D.Faddeev,
\newblock{Quantization of Lie Groups and Lie Algebras,}
\newblock{\it Leningrad.Math.Journal,} {\bf  1}, 193 (1990).




\bibitem{RT}
N.Yu.Reshetikhin, V.G.Turaev,
\newblock{Ribbon Graphs and their invariant derived from quantum groups,}
\newblock{\it Comm.Math.Phys,} {\bf 127 }, (1990).


\bibitem{Ru}
B.Ye.Rusakov,
\newblock{Loop averages and Partition functions in U(N) gauge theory on
two-
dimensional Manifolds,}
\newblock{\it Mod.Phys.Lett.A,} {\bf 5}, 693 (1990).

\bibitem{Wi}
E.Witten,
\newblock{On quantum gauge theories in two dimensions,}
\newblock{\it Comm.Math.Phys,} {\bf 141}, 153 (1991).

\bibitem{Wo}
S.L.Woronowicz,
\newblock{Compact Matrix Pseudogroups,}
\newblock{\it Comm.Math.Phys,} {\bf 111}, 613 (1987).

\end{thebibliography}

\end{document}